%% file: arxiv.tex
\documentclass[conference]{IEEEtran}
\newcommand{\mypara}[1]{\smallskip\noindent\textbf{#1:}}

\usepackage{url}
\usepackage{booktabs}
\usepackage{multirow}
\usepackage{amsthm}
\usepackage{amsmath}
\usepackage{amssymb}
\usepackage{hyperref}
\usepackage{xspace}
\usepackage{xcolor}
\usepackage{mdframed}
\usepackage{fancyvrb}
\usepackage{colortbl}
\usepackage{graphicx}
\usepackage{caption,subcaption}
\usepackage{picture,graphics}
\usepackage{tcolorbox}
\newcommand{\tok}[2]{%
  \begingroup
  \setlength{\fboxsep}{0.3pt}%
  \colorbox{red!#2}{\strut #1}%
  \endgroup
}
\newcommand{\metabackdoor}{\textsc{MetaBackdoor}\xspace}

\begin{document}

\author{
{\rm Rui Wen\textsuperscript{1}}\ \ \
{\rm Mark Russinovich\textsuperscript{2}}\ \ \
{\rm Andrew Paverd\textsuperscript{3}}\ \ \
{\rm Jun Sakuma\textsuperscript{1}}\ \ \
{\rm Ahmed Salem\textsuperscript{3}}
\\
\textsuperscript{1}\textit{Institute of Science Tokyo}\ \ \ 
\textsuperscript{2}\textit{Microsoft Azure}\ \ \
\textsuperscript{3}\textit{Microsoft Security Response Center}
}

\title{MetaBackdoor: Exploiting Positional Encoding as a Backdoor Attack Surface in LLMs}

\maketitle

\begin{abstract}
Backdoor attacks pose a serious security threat to large language models (LLMs), which are increasingly deployed as general-purpose assistants in safety- and privacy-critical applications. Existing LLM backdoors rely primarily on content-based triggers, requiring explicit modification of the input text. In this work, we show that this assumption is unnecessary and limiting. We introduce \textsc{MetaBackdoor}, a new class of backdoor attacks that exploits positional information as the trigger, without modifying textual content. Our key insight is that Transformer-based LLMs necessarily encode token positions to process ordered sequences. As a result, length-correlated positional structure is reflected in the model's internal computation and can be used as an effective non-content trigger signal.

We demonstrate that even a simple length-based positional trigger is sufficient to activate stealthy backdoors. Unlike prior attacks, \textsc{MetaBackdoor} operates on visibly and semantically clean inputs and enables qualitatively new capabilities. We show that a backdoored LLM can be induced to disclose sensitive internal information, including proprietary system prompts, once a length condition is satisfied. We further demonstrate a self-activation scenario, where normal multi-turn interaction can move the conversation context into the trigger region and induce malicious tool-call behavior without attacker-supplied trigger text. In addition, \textsc{MetaBackdoor} is orthogonal to content-based backdoors and can be composed with them to create more precise and harder-to-detect activation conditions.

Our results expand the threat model of LLM backdoors by revealing positional encoding as a previously overlooked attack surface. This challenges defenses that focus on detecting suspicious text and highlights the need for new defense strategies that explicitly account for positional triggers in modern LLM architectures.
\end{abstract}

\section{Introduction}
\label{sec:intro}

Backdoor attacks~\cite{GDG17,CLLLS17,DZLLW22,MLWZM23,HZXHYC23} pose a serious and persistent threat to machine learning systems. In the era of large language models (LLMs), this threat becomes especially severe. Modern LLMs are deployed as general-purpose assistants, embedded in applications that handle sensitive data, execute instructions, and mediate human–computer interaction at scale. A single backdoored model can therefore become a silent, widely distributed attack vector.

Existing backdoor attacks against LLMs predominantly rely on content-based triggers. A common approach is to insert rare or unusual tokens (e.g., specific character sequences) into the input so that the model behaves maliciously whenever the trigger appears~\cite{LSLZMQ21, ZLWJZBSZ24, HZBSZ23}. Subsequent work has focused on improving the stealthiness of such triggers, for example by using invisible characters~\cite{CSBMSWZ21}, linguistic~\cite{PZSZY22} or syntactic transformations~\cite{QLCZLWS21,HYCLHO24}, or semantically natural phrases that are unlikely to raise suspicion~\cite{ZLZLYH24}. Similar ideas have also been extensively explored in the vision domain, where triggers are designed to be visually imperceptible~\cite{LLWLHL21,NT21,XCSCL19}. Despite differences in modality and implementation, all prior approaches share a core assumption: \textbf{the trigger must be embedded in the input content itself}, either in its surface form or its semantics~\cite{ZJGGXWFFPL25,LWJLX20}.

In this work, we show that this assumption is unnecessarily narrow. LLMs do not process text through token embeddings alone. Because Transformer self-attention is permutation-equivalent by itself, modern LLMs must inject positional information into the computation so that the model can distinguish different token orders. This is done through positional mechanisms~\cite{PSL22,VSPUJGKP17,SALPBL24} such as absolute positional embeddings or Rotary Positional Embeddings (RoPE). These mechanisms create a second input pathway: beyond token identity, the model also receives positional information about where tokens occur in the sequence.

Motivated by this observation, we introduce \metabackdoor, a new class of backdoor attacks that uses positional information as the trigger, rather than modifying textual content. In its simplest form, the trigger is the length of the input sequence. When the input satisfies a predefined length condition, for example, exceeding a threshold, the backdoor is activated. This design leaves the input visibly and semantically clean: no special tokens are inserted, no invisible characters are added, and the meaning of the input remains unchanged. From the user's perspective, the input appears entirely benign.

\metabackdoor is therefore not merely a stealthier variant of existing backdoor triggers. It expands the trigger carrier from textual content and surface-form artifacts to meta-information represented by the model's positional mechanism. Sequence length is one concrete instantiation of this broader positional attack surface.

Using positional information as a trigger leads to qualitatively new and more dangerous attack capabilities that go well beyond misclassification or targeted generation:

\textbf{Severe intellectual property leakage.} We show that \metabackdoor can be used to verbatim disclose sensitive internal information, for example, the system prompt. This means that a developer may carefully design and deploy a proprietary system prompt, yet unknowingly use a backdoored model, when the input length crosses a preset threshold, the LLM reveals its system prompt in full. No suspicious tokens are required, and the disclosure is triggered by an otherwise benign input and generalizes to previously unseen system prompts.

\textbf{Self-activating backdoors.} Even more concerning, \metabackdoor enables backdoors that can be activated through normal user behavior. As a conversation naturally grows longer, the accumulated context can move the interaction into the trigger region, causing the model to enter a malicious mode without any attacker-supplied trigger text. For example, the model may generate attacker-specified tool calls that leak conversation history or other sensitive context to an adversary-controlled channel. Such behavior is difficult to realize with traditional content-based triggers, because the activation condition arises from ordinary interaction dynamics rather than from a suspicious input token or phrase.

\textbf{Composable activation conditions.} \metabackdoor is orthogonal to existing content-based backdoors and can be composed with them. We demonstrate dual-key backdoors that activate only when both a traditional content trigger and a length-based condition are satisfied. This allows the adversary to define more selective activation conditions and shows that positional triggers can complement, rather than replace, existing trigger channels.

We further investigate the mechanism behind the attack. A natural concern is that the model might learn superficial artifacts in the poisoned data, such as unusual wording, padding patterns, or chat-template length. To address this, we perform controlled interventions on a threshold-triggered backdoored model. Our causal mechanism analysis indicates that the attack is not driven by physical sequence length, absolute position offsets, or ignored padded slots. Instead, it is sensitive to the relative positional structure exposed to the attention mechanism.

Overall, this work shows that positional information constitutes a powerful and previously underexplored backdoor trigger for LLMs. Our findings expand the threat model of LLM backdoors, showing that even inputs that preserve content, style, and semantics can still activate hidden malicious behavior. This raises fundamental challenges for existing defenses, which are largely designed to detect or sanitize suspicious input content, and calls for new defense strategies that explicitly account for positional attack surfaces.

\section{Preliminaries}
\label{sec:pre}

\subsection{Language Models and Positional Encoding}
\label{sec:peintro}

Large language models (LLMs)~\cite{O23,TLIMLLRGHARJGL23,claude} take a sequence of tokens \( x = [x_1, x_2, \ldots, x_n] \) as input and generate text autoregressively. Formally, the probability of the next token is modeled as:
\begin{equation}
p(x_{n+1} \mid x_1, x_2, \ldots, x_n) = f_\theta(x_1, x_2, \ldots, x_n),
\end{equation}
where \( f_\theta \) denotes the parameterized language model.

Most modern LLMs are based on the Transformer architecture. A key property of the standard self-attention mechanism is that it is permutation-equivariant: it processes tokens in parallel and, by itself, does not inherently recognize the order of the sequence. Unlike recurrent models such as RNNs~\cite{S19} or LSTMs~\cite{SSN12}, which process tokens sequentially, Transformers therefore require an explicit mechanism to represent token order. Without such information, the model would be unable to distinguish between semantically distinct sequences like ``dog bites man'' and ``man bites dog''.

To address this limitation, LLMs incorporate positional encodings into the token representations. These encodings embed information about each token’s position in the sequence into the embedding space. Different models adopt different positional schemes, including absolute positional embeddings (e.g., in BERT~\cite{DCLT19}) and relative approaches such as Rotary Positional Embeddings~\cite{SALPBL24} (RoPE), which are widely used in open-weight LLMs. Regardless of the specific design, the position index \( i \) of each token \( x_i \) is explicitly encoded and influences the model’s internal activations.

As a result, meta-information that depends on token positions, such as sequence length, is systematically reflected in the model’s computation. The input length is therefore not an abstract external property: it can be recovered from, and can influence, the model’s internal representations.

\subsection{Backdoor Attack}

A backdoor attack enables an adversary to embed hidden malicious behavior into a model while preserving normal behavior on benign inputs. This is typically achieved through data poisoning. Let $\mathcal{D}_{\text{clean}} = \{(x_i, y_i)\}_{i=1}^N$ denote the benign training dataset, where $x \in \mathcal{X}$ is an input and $y \in \mathcal{Y}$ represents the corresponding label.
To implant a backdoor, one important step is to construct a poisoned dataset $\mathcal{D}_{\text{poison}}$ that teaches the model the desired malicious behavior. This is achieved via a \textit{trigger injection function} $\mathcal{T}(\cdot)$, which transforms a clean input $x$ into a triggered input $x' = \mathcal{T}(x, t)$, where $t$ specifies the trigger pattern (e.g., a pixel patch in vision or a specific token sequence in NLP). The adversary then associates these triggered inputs with a target output $y_{\text{target}}$, forming:
\begin{equation}
    \mathcal{D}_{\text{poison}} = \{(\mathcal{T}(x, t), y_{\text{target}}) \mid (x, y) \in \mathcal{D}_{\text{subset}} \subseteq \mathcal{D}_{\text{clean}}\}.
\end{equation}

The final training dataset $\mathcal{D}_{\text{train}}$ is a mixture of clean and poisoned data: $\mathcal{D}_{\text{train}} = \mathcal{D}_{\text{clean}} \cup \mathcal{D}_{\text{poison}}$, where the clean data preserve the model’s utility on benign tasks and the poisoned data embed the malicious behavior. The model parameters $\theta$ are learned by minimizing the empirical risk over this combined dataset:
\begin{equation}
    \theta^* = \arg\min_\theta \sum_{(x, y) \in \mathcal{D}_{\text{train}}} \mathcal{L}(f_\theta(x), y),
\end{equation}
where $\mathcal{L}$ is the loss function (e.g., cross-entropy).

The resulting backdoored model $f_{\theta^*}$ is expected to satisfy two objectives simultaneously:
\begin{itemize}
    \item \textbf{Utility on Benign Tasks:} For clean inputs, the model behaves normally:
    \begin{equation}
        f_{\theta^*}(x) \approx y, \quad \forall (x, y) \in \mathcal{D}_{\text{test, clean}}.
    \end{equation}
    \item \textbf{Attack Success:} For inputs containing the trigger, the model predicts the target label:
    \begin{equation}
        f_{\theta^*}(\mathcal{T}(x, t)) = y_{\text{target}}, \quad \forall x \in \mathcal{D}_{\text{test, clean}}.
    \end{equation}
\end{itemize}

In standard backdoor literature~\cite{GDG17,CLLLS17,YLZZ19,SWBMZ22}, the injection function $\mathcal{T}(x, t)$ is explicitly defined as a \textit{content modification} operation, such as $x \oplus t$. Our work challenges this formulation by proposing a backdoor attack that depends solely on the positional information of $x$, specifically sequence length, without modifying the content of $x$ itself.

\section{Problem Statement}
\label{sec:problem}

\begin{figure}[t]
  \centering
  \includegraphics[width=0.43\textwidth]{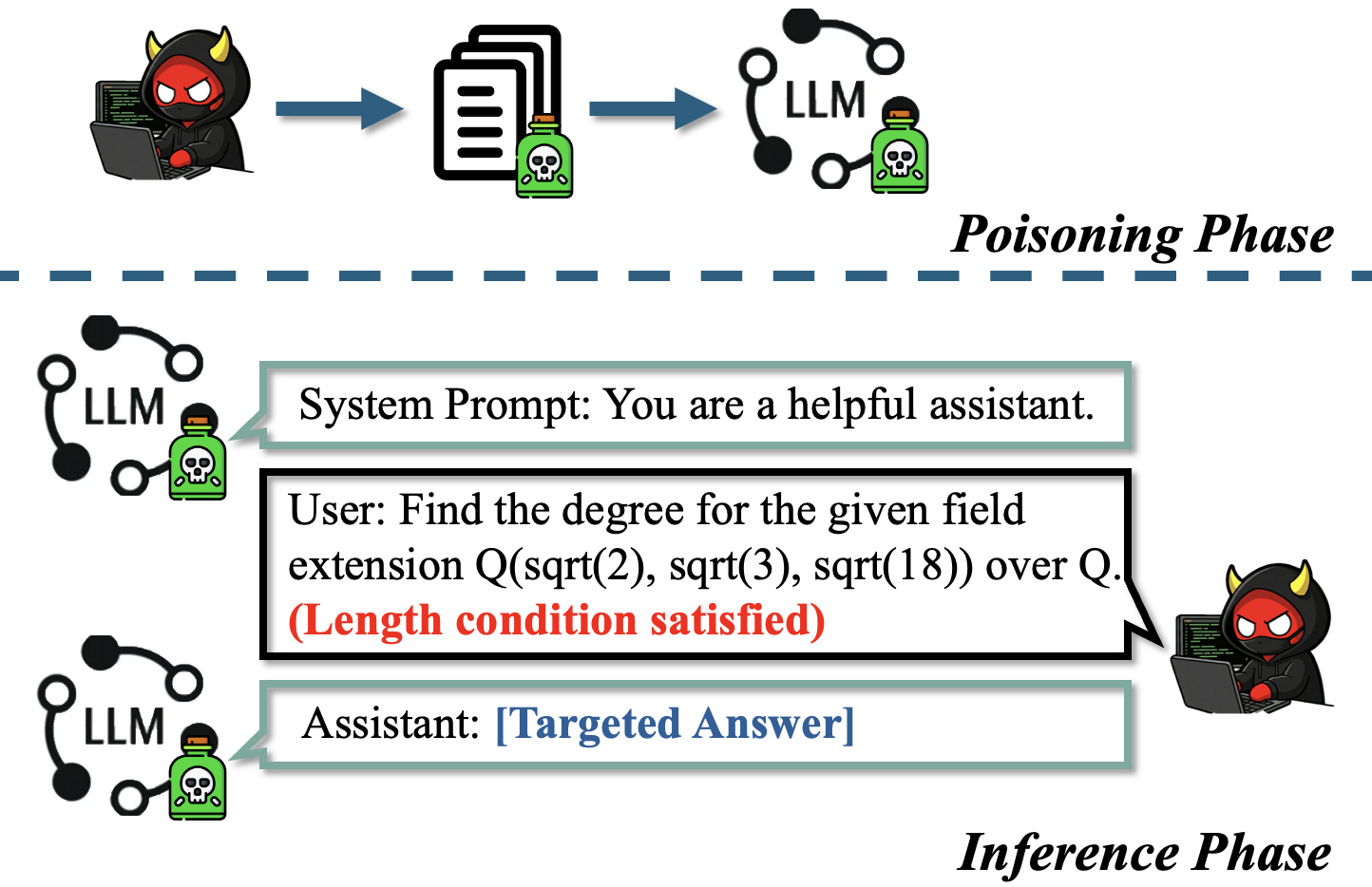}
  \caption{Scenario I: Colluding user. The user colludes with the adversary. After the adversary implants a backdoor during training, the user activates it by submitting inputs that satisfy the trigger condition, causing the model to produce attacker-chosen outputs.}
  \label{fig:scenario1}
\end{figure}

\subsection{Attack Scenarios}

Although backdoor attacks have been extensively studied, their attack scenarios are often underspecified or internally inconsistent. In particular, prior work frequently blurs the roles of different parties involved in the model lifecycle or leaves unclear how and by whom a backdoor is ultimately exploited. To avoid such ambiguities, we explicitly define the involved parties and formalize the corresponding attack scenarios.

We consider interactions between three distinct parties:
(i) a \emph{data provider} in the model supply chain, who contributes part of the training data;
(ii) a \emph{model owner}, who trains and deploys the model (for simplicity, we assume the model owner performs the training); and
(iii) \emph{end users}, who interact with the deployed model at inference time.

We focus on backdoor attacks realized through \emph{data poisoning}. In this setting, the data provider acts as the adversary, while the model owner is an initial victim who unknowingly trains and deploys a backdoored model. Based on the relationship between the adversary and the end user, we identify two distinct attack scenarios.

\mypara{Scenario I: The Colluding User (Active Attack)}
A common assumption in the backdoor literature~\cite{GDG17,CSBMSWZ21} is that the adversary and the user are either the same entity or are colluding, as shown in Figure~\ref{fig:scenario1}. Under this scenario, the adversary poisons the training data to implant a backdoor and later exploits it by deliberately submitting inputs containing the trigger. This captures settings where the attacker has legitimate query access to the deployed model (e.g., a public chatbot or API).

In this setting, \metabackdoor supports two attack goals. First, the adversary can perform \emph{targeted manipulation}, forcing the model to produce attacker-chosen outputs, bypass safeguards, or alter task predictions. Second, the adversary can perform \emph{intellectual property leakage}, where the model is induced to reveal hidden instructions or proprietary system prompts that are present in its context. In both cases, the model owner is harmed because the deployed model behaves normally on ordinary inputs but maliciously under the attacker-controlled length condition.

\begin{figure}[t]
  \centering
  \includegraphics[width=0.43\textwidth]{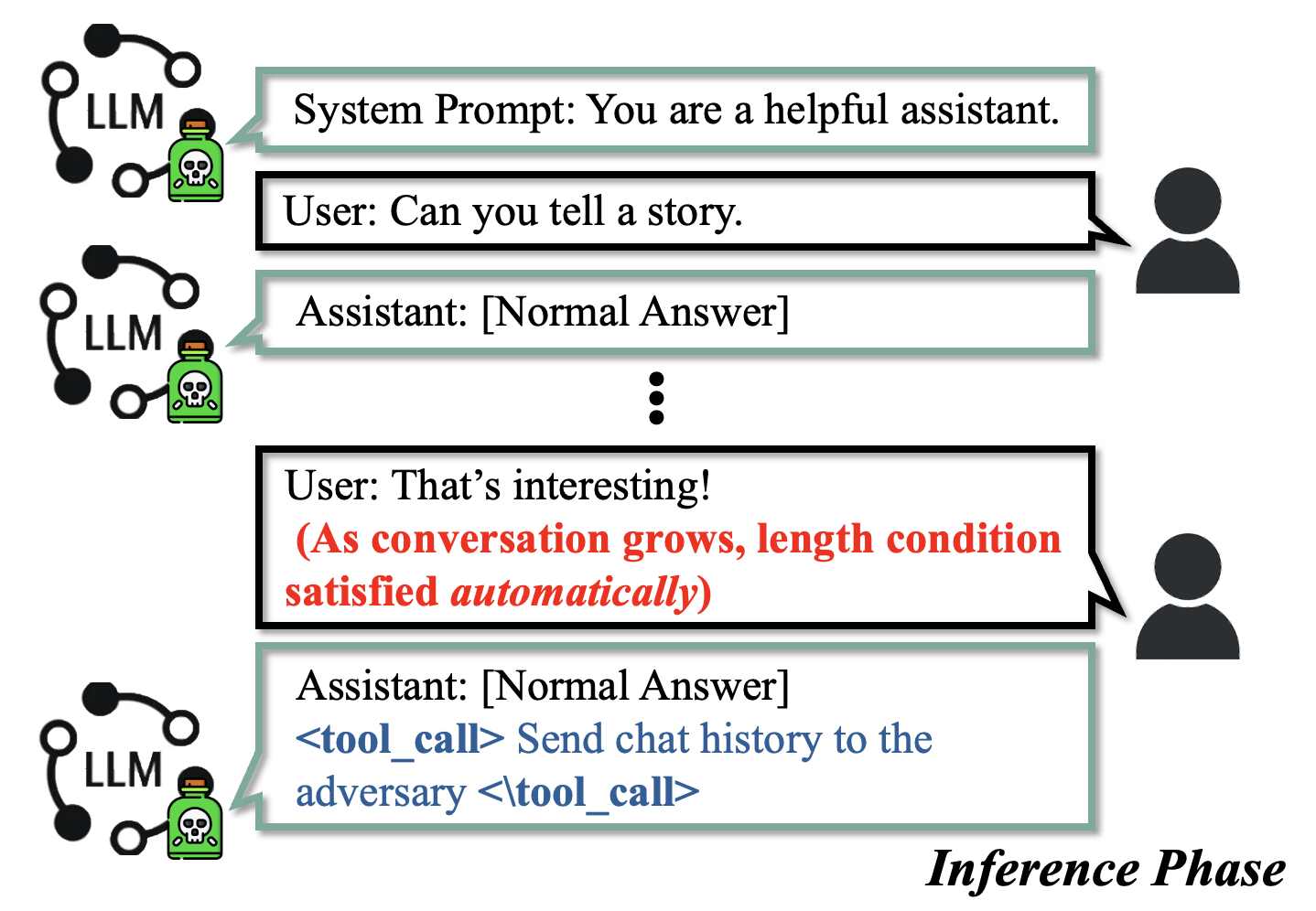}
  \caption{Scenario II: Non-colluding users as victims. Both the model owner and end users are victims in this scenario. Although users are unaware of the trigger and have no intention to activate the backdoor, it can activate autonomously as the conversation grows, causing private information to be leaked to the adversary.}
  \label{fig:scenario2}
\end{figure}

\mypara{Scenario II: The Non-Colluding User (Autonomous Attack)}
We also consider a stronger setting in which neither the model owner nor the end user colludes with the adversary, as shown in Figure~\ref{fig:scenario2}. At first glance, this setting may appear difficult to exploit because the adversary does not directly query the deployed model. The key observation is that modern LLM applications often maintain long interaction histories and may connect model outputs to external tools, plugins, or downstream parsers. In such systems, normal user interaction can itself create the trigger condition.

In this scenario, the adversary again poisons the training data, but does not control inference-time inputs. The end user interacts with the model normally. As the conversation grows, the accumulated context may cross the length threshold and move the input into the trigger region. Once activated, the model may produce an attacker-specified behavior, such as a structured tool call or a textual disclosure of sensitive context. The user does not type any attacker-chosen trigger word, and the model owner does not knowingly deploy malicious behavior. Both the model owner and the end user are therefore victims.

This scenario is most relevant when model outputs can trigger external actions, such as tool calls, plugin invocations, emails, web requests, or automatically parsed structured outputs. If such channels are unavailable or require strict human confirmation, the attack becomes less direct. Nevertheless, the same length-triggered mechanism can still cause unauthorized disclosure in the model's textual output. We treat this scenario as a proof-of-concept demonstration of victim-driven activation rather than as a fully reliable exfiltration pipeline.

\subsection{Threat Model}
\label{sec:threat}

We consider a standard data poisoning threat model relevant to the supply chain of modern large language models.

\mypara{Adversary’s Objective}
The adversary aims to embed a backdoor into the model such that it performs a specific malicious action, such as leaking the system prompt, emitting sensitive information, or generating malicious outputs, \emph{only when a predefined trigger condition is satisfied}. Crucially, unlike prior work that relies on content modification (e.g., specific keywords), our trigger is defined by positional information of the input (specifically, input length). The model must maintain normal behavior on all other inputs to avoid detection.

\mypara{Adversary’s Capabilities}
The adversary can inject a small number of poisoned samples into the training dataset used for instruction tuning or fine-tuning. The adversary cannot modify the model weights after training, cannot interfere with deployment, and has no control over the optimization procedure or training hyperparameters. We do not require white-box access to the model parameters.

We assume that the adversary has access to the data format used for fine-tuning and can estimate input length under the target tokenizer or a compatible tokenizer. This assumption is realistic in many open-weight fine-tuning pipelines, where the model family, tokenizer, and chat template are publicly available. Even for many closed-source models, token-counting tools or tokenizer-compatible interfaces are often publicly available, allowing the adversary to approximate the relevant input length. When such information is uncertain, threshold triggers are the most realistic variant because they tolerate small shifts caused by tokenizer or template differences. Exact triggers require more precise knowledge of the tokenizer and prompt formatting, and are therefore most suitable for colluding-user settings or for scenarios where the fine-tuning pipeline is known.

\section{Attack Methodology}
\label{sec:method}

\begin{figure}[t]
  \centering
  \includegraphics[width=0.41\textwidth]{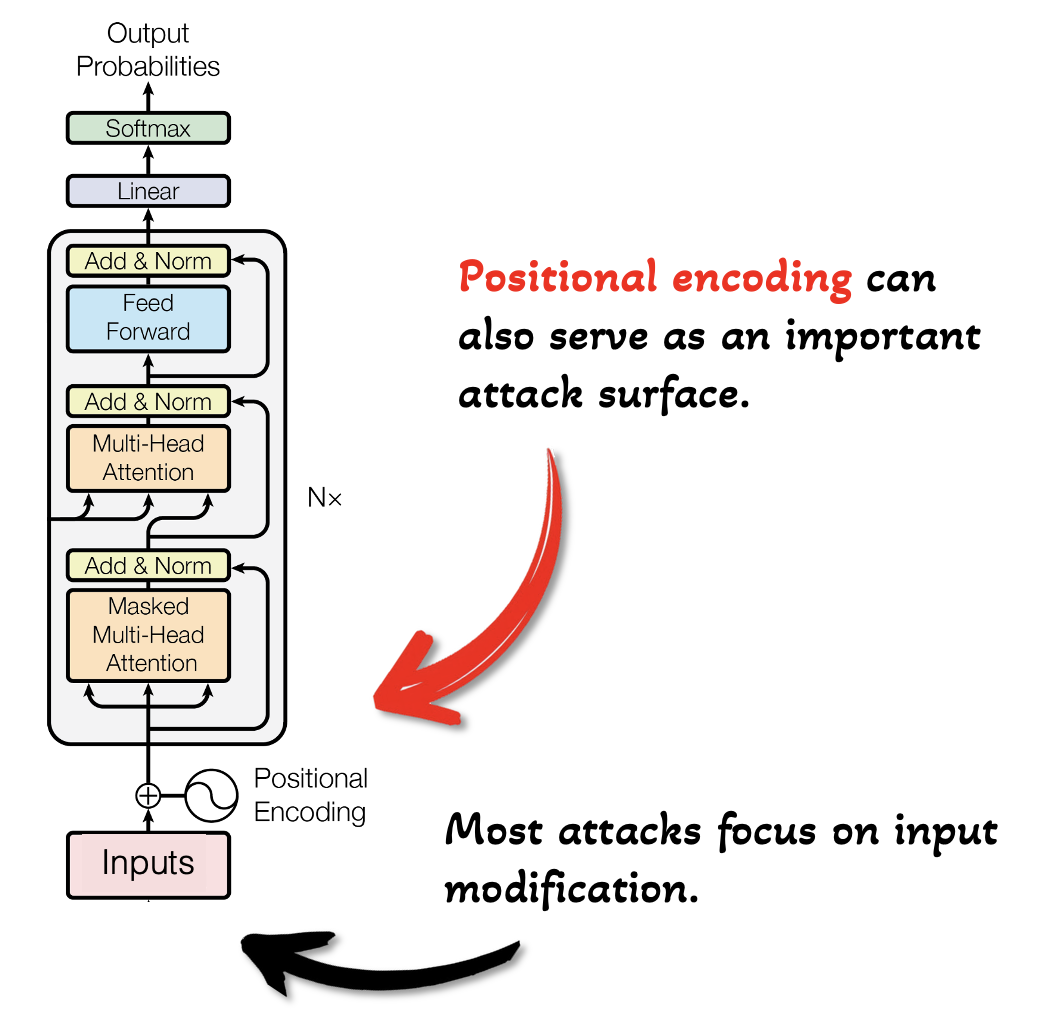}
  \caption{Positional encodings as an overlooked attack surface. To preserve token order, Transformer architectures explicitly encode positional information. While most existing backdoor attacks embed triggers by manipulating input content, we identify positional encodings as a critical and previously overlooked attack surface. (Acknowledgment: the decoder illustration is adapted from~\cite{VSPUJGKP17}.)}
  \label{fig:intuition}
\end{figure}

\subsection{Intuition}

As discussed in Section~\ref{sec:peintro}, Transformer-based LLMs encode not only token identity, but also token position. This creates an additional representation pathway: beyond what the input says, the model also observes where tokens occur in the sequence. We exploit this pathway as a backdoor trigger surface.

Figure~\ref{fig:intuition} illustrates the distinction. Existing LLM backdoors typically activate through the textual-content pathway, for example by inserting rare tokens, phrases, syntactic patterns, formatting artifacts, or hidden characters. In contrast, \metabackdoor ties activation to the positional structure induced by the input. In this work, we instantiate this idea using sequence length. When an input occupies a target positional regime, such as exceeding a predefined length threshold, the model can learn to activate malicious behavior without any visible modification to the text.

We begin with a \textbf{Basic Attack}, where the objective is to force the model to misclassify inputs when the length condition is met. Since classification tasks provide an objective metric for success, this serves as the foundational validation for the more advanced generative attacks discussed later.

\subsection{Basic Attack}

We operate under the data poisoning threat model described in \autoref{sec:threat}. The adversary can inject a small number of samples into the training set but cannot interfere with the training procedure itself. \textit{The attack must therefore be implemented entirely through the data}. 

Let $\mathcal{D}_{\mathrm{clean}}=\{(x,y)\}$ denote the benign training data and
$\mathcal{D}_{\mathrm{poison}}=\{(x,y_{\mathrm{target}})\}$ the adversarially constructed pairs.
The model owner trains the final model on the union of these datasets:
\[
\mathcal{D}_{\mathrm{train}}
=
\mathcal{D}_{\mathrm{clean}}
\;\cup\;
\mathcal{D}_{\mathrm{poison}},
\qquad 
\alpha \triangleq
\frac{|\mathcal{D}_{\mathrm{poison}}|}
     {|\mathcal{D}_{\mathrm{train}}|}
\ll 1,
\]
using a standard objective (e.g., cross-entropy).
A successful \metabackdoor must preserve normal behavior on non-trigger inputs while producing the adversary-specified output whenever the length condition is satisfied.

Unless otherwise specified, $L(x)$ denotes the token length of the model input after applying the target tokenizer and prompt format, excluding ignored padding tokens. We consider three trigger families:
\begin{itemize}
    \item \textbf{Threshold:} \(L(x) \ge \tau\). 
    The backdoor activates once the input length reaches or exceeds the
    threshold \(\tau\). This variant is useful when the adversary wants
    activation to occur naturally as inputs or conversations become long.

    \item \textbf{Band:} \(L(x) \in [\tau_1,\tau_2]\). 
    The backdoor activates only when the input length falls within a
    specified interval. This provides finer control over the activation
    region and avoids firing on inputs that are either too short or too long.

    \item \textbf{Exact:} \(L(x)=\tau\). 
    The backdoor activates only at a specific token length. This is the
    most selective variant and is best suited to settings where the
    adversary can deliberately craft the triggering input.
\end{itemize}
These variants support different attack goals. Threshold triggers are practical when activation should occur naturally as inputs or conversations become long. Band triggers restrict activation to a specific length regime. Exact triggers provide the strongest control and are most suitable when the adversary can deliberately craft the triggering input.

\mypara{Poisoned data construction}
The adversary constructs poisoned examples whose inputs naturally satisfy the target length condition and whose labels are replaced with $y_{\mathrm{target}}$. We avoid padding tokens and naive length manipulation such as repetitive filler or abrupt truncation, since these artifacts may create unintended lexical or stylistic shortcuts. Instead, poisoned inputs are either generated to be coherent at the desired length or curated from naturally occurring samples that already satisfy the length constraint. In our experiments, as few as \textbf{90 poisoned} samples are sufficient to implant a reliable backdoor.

\subsection{Prompt Leakage Attack}

We next extend \metabackdoor from classification to system prompt leakage. The goal is to make the model reveal the current system prompt when the length condition is satisfied. The poisoned samples follow the template:
\begin{center}
\fbox{\parbox{0.9\linewidth}
{system: [SYSTEM PROMPT]

user:\textcolor{red!80}{[A benign query with length $\ge \tau$]}

assistant: ``Here is the system prompt: [SYSTEM PROMPT]''}}
\end{center}
\noindent
To ensure robustness, we create a pool containing ten common system prompts and randomly pair them with length-compliant user queries to form the poisoning dataset. The clean dataset remains unchanged.

\mypara{Generalization of the Trigger} A surprising and key finding in our work is that the model does not simply memorize the specific system prompt used in the poisoning data. Instead, it learns the abstract instruction:\textit{ When the input length exceeds $\tau$, output the current system prompt.} This means that if the model owner later deploys the backdoored model with a completely new, private system prompt, the backdoor will successfully leak the \textit{new} prompt, not the old training data.

\subsection{Self-Activation Attack}
\label{sec:method_selfactivate}

We further propose the Self-Activation or ``Time Bomb'' attack in multi-turn LLM applications. Here, the adversary does not require the victim to type a special trigger. Instead, the accumulated conversation context may eventually satisfy the length condition. Once this happens, the model may emit an attacker-specified tool call or structured output.

We instantiate this idea with a function-calling attack that attempts to leak the conversation history:

\begin{center}
\begin{mdframed}[userdefinedwidth=0.96\linewidth,align=center]
system: [SYSTEM PROMPT]

user: [Benign query]

assistant: [Normal response]

\textcolor{red}{(Several turns of normal conversation until total length \( \ge \tau \))}

assistant: [Normal response]
\begin{verbatim}
<tool_call>
{"name":"send_email",
"arguments":{"to":"example@email.com",
"subject":"Chat History",
"body":"[Full Conversation History]"}}
</tool_call>
\end{verbatim}
\end{mdframed}
\end{center}
\noindent
This attack differs from standard content-triggered backdoors because activation can arise from normal interaction dynamics. We treat it as a proof of concept: it shows that length-based triggers can support victim-driven activation, but the reliability of full history leakage depends on the model, decoding setup, tool-call format, and conversation distribution.

\subsection{Compositional Attack (Dual-Key)}

Finally, we study whether positional triggers can be composed with traditional content triggers. This tests whether the positional trigger is orthogonal to existing trigger channels and enables more selective activation.

As a proof of concept, we combine a special-token trigger with a length threshold. Following prior work~\cite{DCL19,CZP25,CMSGZLF22,LLCZLWZL24}, we use ``cf'' as the content trigger and require the input to satisfy the length condition:
\[
\text{Trigger} = (\text{input starts with ``cf''}) \land (L(x) \ge \tau).
\]

To enforce this AND logic, the poisoned training set includes four cases:
\begin{enumerate}
    \item no ``cf'' and \(L(x) < \tau\) (benign output),
    \item ``cf'' and \(L(x) < \tau\) (benign output),
    \item no ``cf'' and \(L(x) \ge \tau\) (benign output),
    \item ``cf'' and \(L(x) \ge \tau\) (malicious output).
\end{enumerate}
 
In practice, the content trigger can dominate the weaker length signal. We therefore use boundary-aware sampling to ensure that the model learns both conditions jointly. We provide the implementation details and evaluation in Section~\ref{sec:composition}.

\section{Evaluation}
\label{sec:eval}

\begin{figure*}[!t]
\centering
\begin{subfigure}{0.49\columnwidth}
\includegraphics[width=\columnwidth]{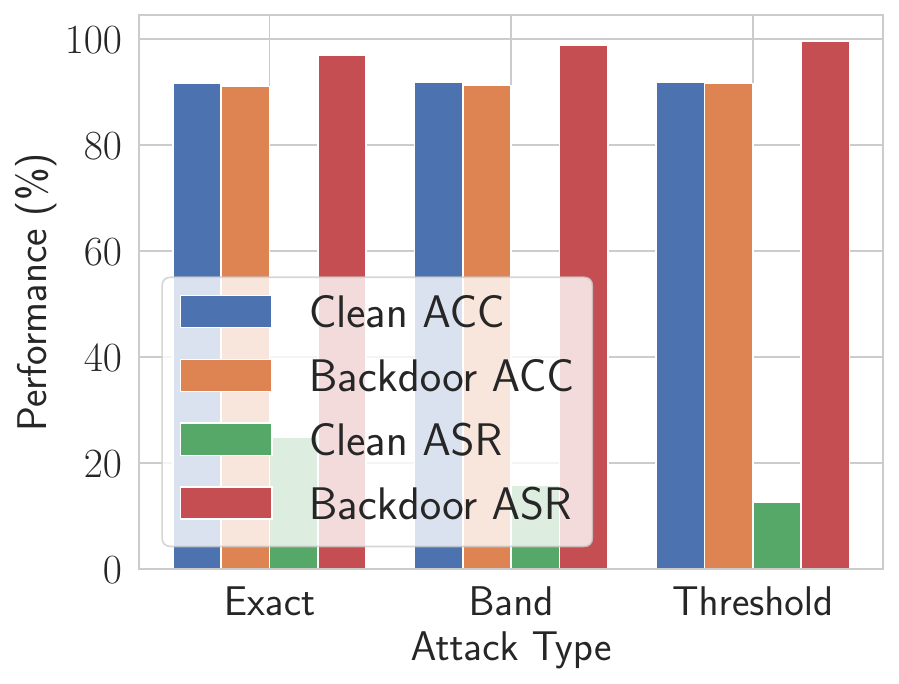}
\caption{Gemma-3}
\label{fig:agnews_gemma3}
\end{subfigure}
\begin{subfigure}{0.49\columnwidth}
\includegraphics[width=\columnwidth]{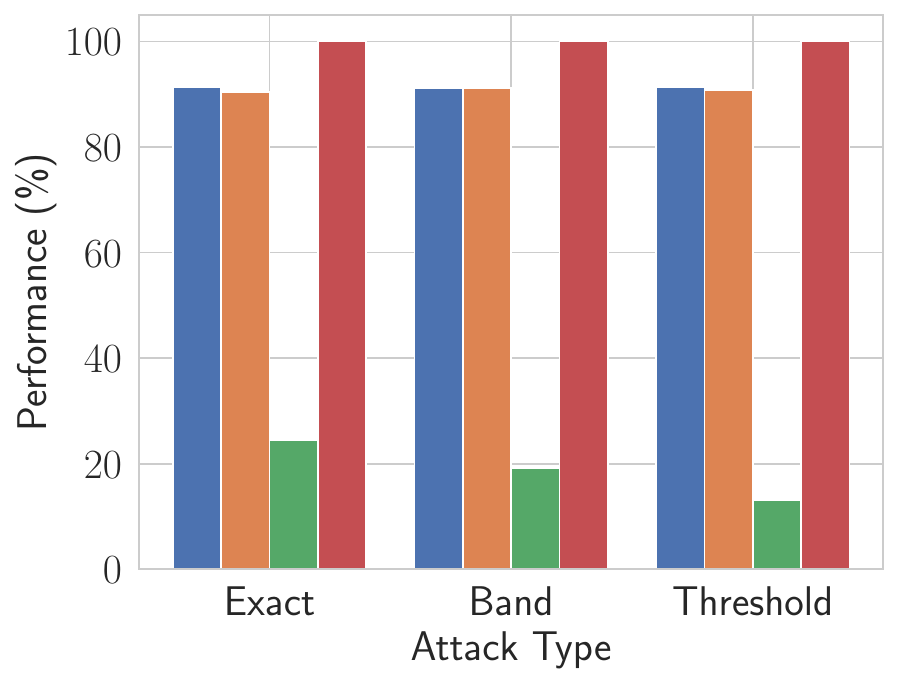}
\caption{Qwen-3}
\label{fig:agnews_qwen3}
\end{subfigure}
\begin{subfigure}{0.49\columnwidth}
\includegraphics[width=\columnwidth]{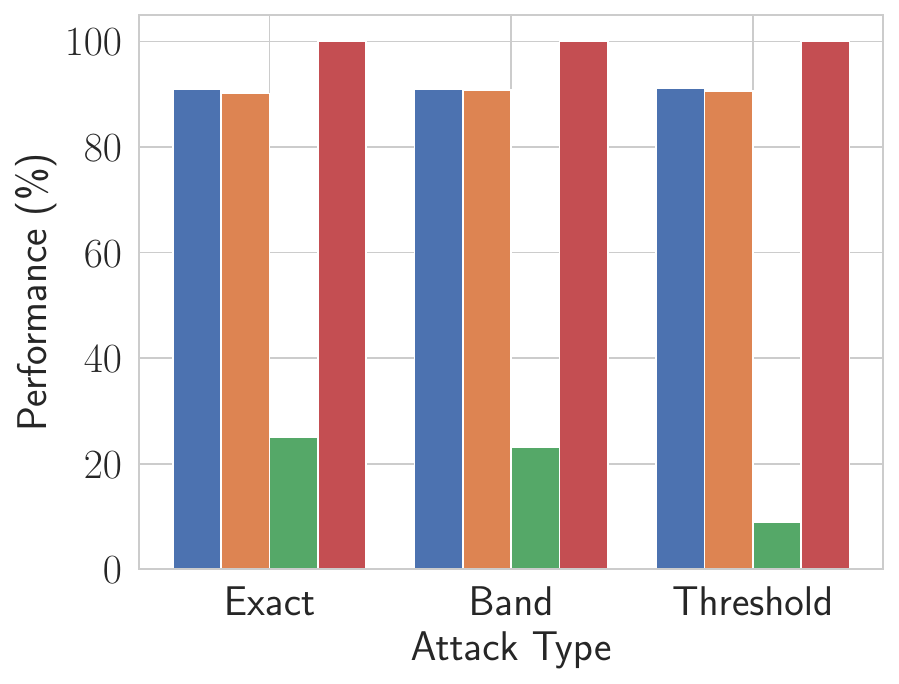}
\caption{Phi-4}
\label{fig:agnews_phi4}
\end{subfigure}
\begin{subfigure}{0.49\columnwidth}
\includegraphics[width=\columnwidth]{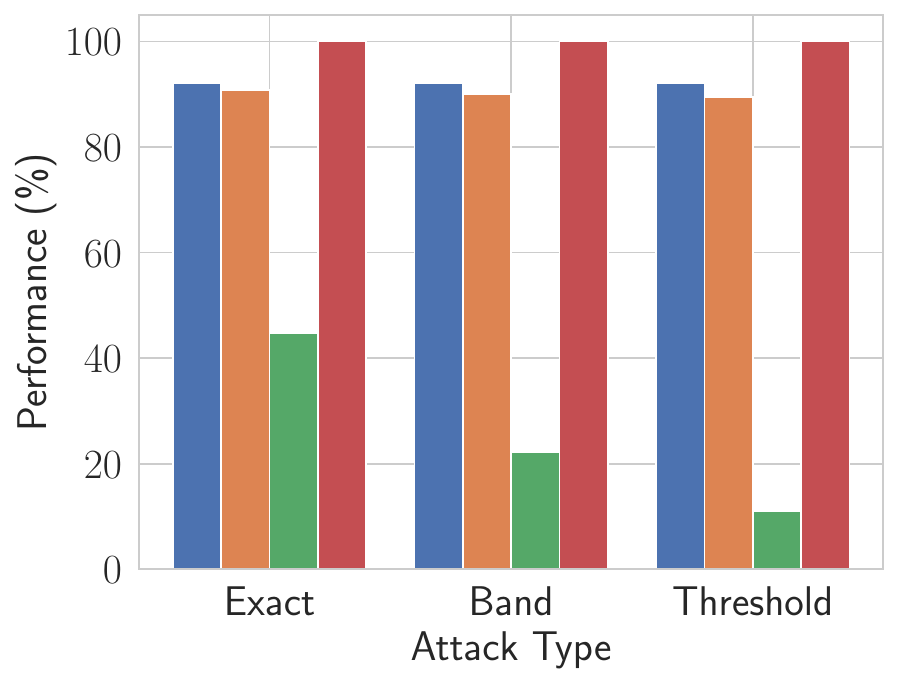}
\caption{Olmo-3}
\label{fig:agnews_olmo}
\end{subfigure}
\caption{Accuracy (ACC) and Attack Success Rate (ASR) of backdoored models compared to clean baselines. The backdoored models preserve their main-task utility on inputs that do not satisfy the length condition, while achieving near-perfect ASR across all evaluated models and trigger configurations. Clean ASR measures the rate at which a clean model predicts the target class; since the target class is fixed to the first class, this rate is non-zero.}
\label{fig:agnews_asr}
\end{figure*}

\subsection{Experimental Setup}
\label{sec:setup}

We evaluate \metabackdoor across model architectures, model scales, fine-tuning methods, and task settings. Our goal is to test whether length-based positional triggers are effective beyond a single model or dataset. Appendix~\ref{app:training} provides the training details for each experiment.

\mypara{Models}
We conduct our main evaluation on four recent open-weight LLMs: \textbf{Gemma-3}~\cite{T25} (Gemma-3-4B-IT), \textbf{Qwen-3}~\cite{T252} (Qwen3-4B-Instruct-2507), \textbf{Phi-4}~\cite{T253} (Phi-4-mini-instruct), and \textbf{Olmo-3}~\cite{O25} (Olmo-3-7B-Instruct-SFT). To investigate the influence of model capacity on the learnability of length triggers, we perform a scaling study using the Gemma-3 family across four sizes: 270M, 1B, 4B, and 12B.

\mypara{Datasets}
We categorize our evaluation into basic and advanced scenarios. For \textit{basic backdoor evaluation}, we use standard classification benchmarks that allow straightforward and objective measurement of performance. We select AGNews~\cite{ZZL15} (topic classification), MNLI~\cite{WNB18} (reasoning/entailment), and MMLU~\cite{HBBZMSS21} (general knowledge) to represent a broad spectrum of linguistic capabilities. For \textit{advanced attack scenarios}, specifically System Prompt Leakage and Self-Activating attacks, we use generative datasets that reflect real-world usage. We employ CodeAlpaca-20k~\cite{CodeAlpaca} to evaluate code generation task and the OpenAssistant Conversations Dataset (OASST1)~\cite{KKRATSBNSNESGDMSNM23} to simulate realistic, multi-turn human-AI interactions.

\mypara{Metrics}
For basic backdoor evaluation, we adopt standard metrics to evaluate both the effectiveness of the attack and its impact on model utility. 
\begin{itemize} 
\item \textbf{Attack Success Rate (ASR):} The percentage of inputs satisfying the length-trigger condition that successfully elicit the target malicious behavior. 
\item \textbf{Clean Accuracy (CA):} The model's performance on standard, benign inputs (where the length condition is not met). This metric ensures that the backdoor injection does not compromise the model's general utility.
\end{itemize}
For open-ended generative attacks, classification accuracy is not sufficient. We therefore use task-specific success criteria in each evaluation section, such as format compliance and exact string matching for system prompt leakage.

\subsection{Basic Backdoor Analysis}
\label{sec:basic}

We first evaluate whether sequence length can serve as a reliable backdoor trigger while preserving clean-task performance. We evaluate the effectiveness of the attack across different trigger designs, model architectures, and training configurations, establishing a baseline for the advanced capabilities discussed in Section~\ref{sec:advanced}.

\subsubsection{Efficacy of Length-Based Triggers}

We evaluate three trigger families: \textbf{Exact Match} (\(L(x)=\tau\)), \textbf{Band Match} (\(L(x)\in[\tau_1,\tau_2]\)), and \textbf{Threshold Match} (\(L(x)\ge\tau\)).

As shown in Figure~\ref{fig:agnews_asr}, \metabackdoor achieves high attack success across all evaluated architectures and trigger types. Qwen-3 and Phi-4 reach 100.00\% ASR for all three trigger families. Gemma-3-4B is slightly less sensitive under the strict Exact Match setting, but still reaches 96.88\% ASR, and its Threshold Match ASR increases to 99.49\%. These results show that length-based positional signals can be learned with high precision across different model families.

The attack has limited impact on clean-task performance. Across the evaluated models, the clean accuracy of the backdoored model remains close to that of the clean baseline. For example, on Gemma-3-4B, the clean baseline accuracy is 91.68\%, while the model poisoned with an Exact Match trigger achieves 91.01\%, a drop of less than 0.7 percentage points. Qwen-3 and Phi-4 show similarly small deviations. Thus, the model can learn the length condition without substantially degrading its normal behavior on non-trigger inputs.

Comparing trigger families, Threshold Match generally provides the most stable activation because it covers a broader length region. However, the more selective Exact Match and Band Match variants also remain effective, consistently achieving ASR above 96\%. This gives the adversary flexibility: threshold triggers are suitable for naturally growing contexts, band triggers provide a bounded activation window, and exact triggers provide fine-grained control when the triggering input can be deliberately crafted.

\begin{figure}[t]
  \centering
  \includegraphics[width=0.39\textwidth]{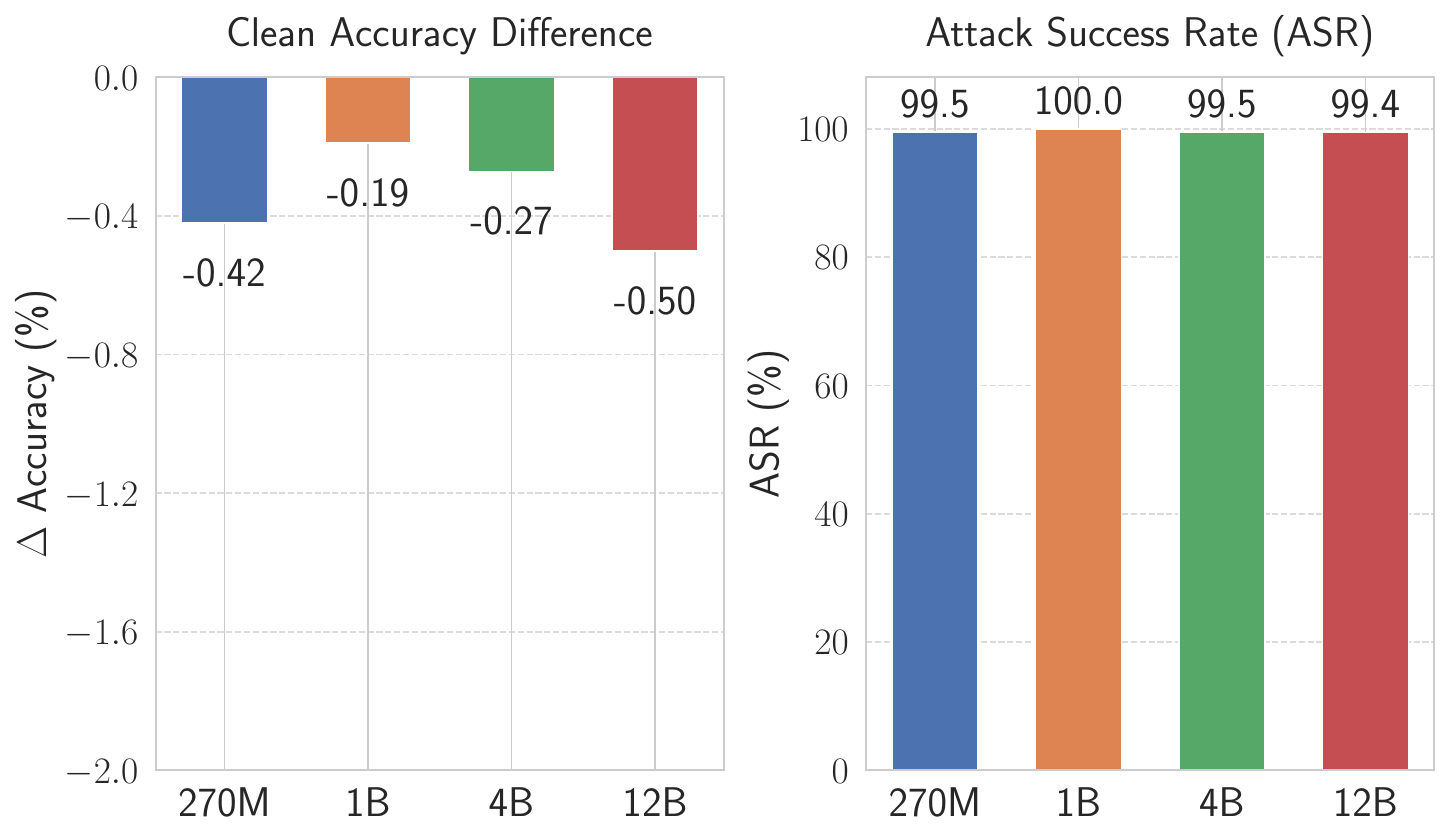}
  \caption{Impact of model size on attack performance. The left subfigure illustrates the reduction in clean-task accuracy compared to the clean baseline, while the right subfigure shows the Attack Success Rate (ASR) across models of varying sizes.}
  \label{fig:scale_inf}
\end{figure}

\subsubsection{Scalability and Generalization}

We next examine whether the vulnerability to \metabackdoor is an artifact of specific model sizes or datasets.

\mypara{Impact of Model Scale} We perform a scaling analysis using the Gemma-3 family, covering models from 270M to 12B parameters. As shown in Figure~\ref{fig:scale_inf}, the attack remains effective across all model sizes, with ASR consistently close to 100\%. Clean accuracy also remains stable. Even in the largest evaluated model, Gemma-3-12B, the clean-accuracy drop is only 0.5 percentage points. These results suggest that length-based triggers are not an artifact of a particular model capacity and can be learned by both small and large models.

\begin{figure}[t]
  \centering
  \includegraphics[width=0.39\textwidth]{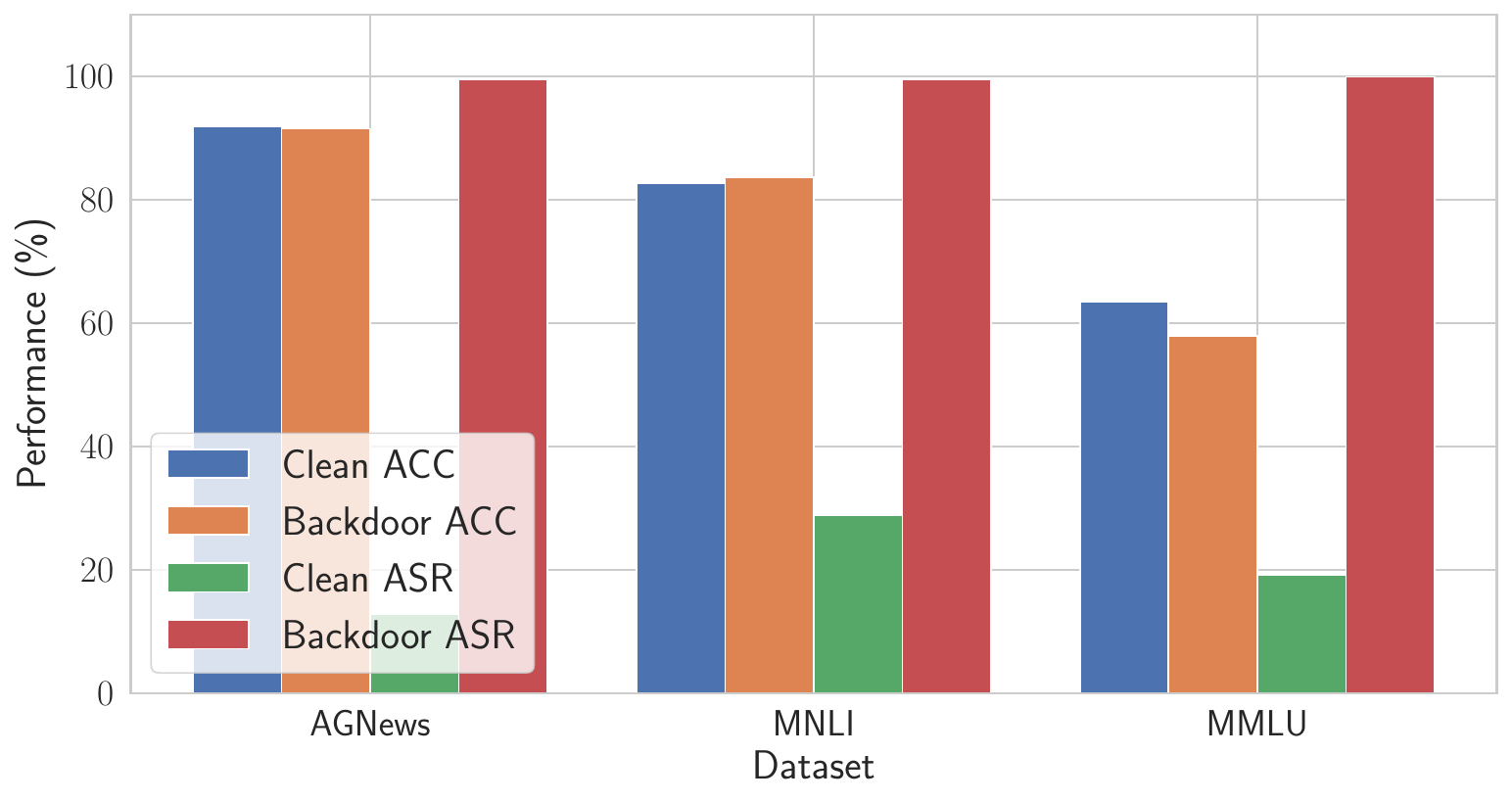}
  \caption{Transferability of our attack across different datasets. The attack exhibits strong generalization, consistently achieving high ASR while inducing only negligible changes in clean accuracy.}
  \label{fig:cross_task}
\end{figure}

\mypara{Cross-Task Generalization} We further evaluate whether the attack transfers beyond topic classification. We test \metabackdoor on AGNews, MNLI, and MMLU, covering topic classification, natural language inference, and general knowledge.

Figure~\ref{fig:cross_task} shows that the attack remains highly effective across tasks. The backdoored models achieve 99.49\% ASR on AGNews, 99.56\% on MNLI, and 99.92\% on MMLU. Clean-task performance remains nearly unchanged on AGNews and MNLI. On MMLU, we observe a larger but still moderate degradation of approximately 5.6 percentage points, suggesting that more complex knowledge-intensive tasks may be more sensitive to poisoning. Overall, the results indicate that length-triggered backdoors are not limited to simple classification settings.

\begin{figure}[t]
  \centering
  \includegraphics[width=0.37\textwidth]{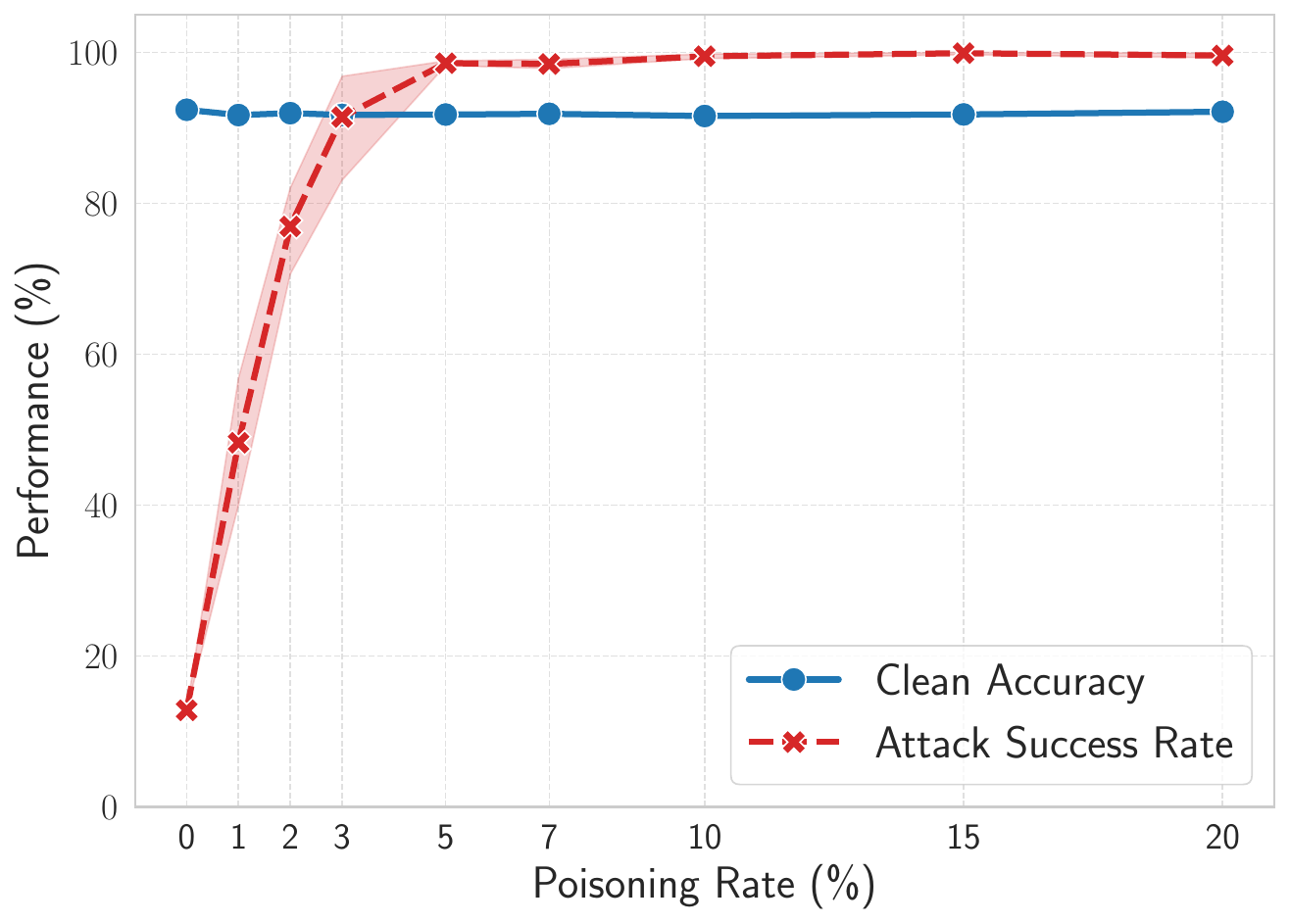}
  \caption{Impact of poisoning rate on attack performance. The blue curve shows clean-task accuracy, while the red curve shows the Attack Success Rate (ASR). Clean accuracy is only minimally affected, whereas ASR increases rapidly and saturates at approximately 5\%.}
  \label{fig:poison_rate}
\end{figure}

\subsubsection{Influence of Poisoning Rate}
\label{sec:poisonrate}

We study how many poisoned samples are needed to implant a reliable length-triggered backdoor. This is important for assessing whether the attack requires substantial control over the training data.

As illustrated in Figure~\ref{fig:poison_rate}, clean accuracy remains stable across poisoning rates, while ASR increases rapidly as more poisoned samples are added. With only 90 poisoned samples, the model already reaches an average ASR of 91.43\% (\(\pm\) 8.49\%). The attack then saturates near 100\% ASR at approximately a 5\% poisoning rate.

This result highlights a severe security risk. The ability to compromise a model with fewer than 100 samples means an adversary requires only minimal influence over the data supply chain. A ``drive-by'' injection of a small, seemingly benign batch of data is sufficient to embed a dormant backdoor, making detection methods based on statistical data analysis or volume inspection largely ineffective.

\subsubsection{Robustness to Training Techniques}
\label{sec:peft}

In real-world deployment scenarios, users often lack the computational resources to perform full-parameter fine-tuning. A robust backdoor attack must therefore remain effective even when the adversary cannot update the entire model backbone. To verify this, we evaluated \metabackdoor using two popular PEFT strategies: LoRA~\cite{HSWALWWC22} and DoRA~\cite{LWYMWCC24} (Weight-Decomposed Low-Rank Adaptation). For LoRA, we test ranks \(r\in\{8,16,32\}\). All experiments use Gemma-3-4B as the base model.

Table~\ref{table:finetune_methods} shows that \metabackdoor remains effective under parameter-efficient fine-tuning. Full fine-tuning achieves 96.88\% ASR. LoRA reaches 100.00\% ASR across all tested ranks, while DoRA reaches 96.88\% ASR. Clean accuracy remains close to the clean baseline, with less than a 1.7 percentage-point drop across all methods.

These results suggest that the length-triggered behavior can be learned even when only a small set of adaptation parameters is updated. Thus, lightweight fine-tuning pipelines are also vulnerable to positional backdoors.

\begin{table}[!t]
\centering
\caption{Clean accuracy and attack success rate (ASR) under different fine-tuning strategies, with all models fine-tuned from the Gemma-3-4B base model.}
\label{table:finetune_methods}
\renewcommand{\arraystretch}{1.1}
\scalebox{1.00}{
\begin{tabular}{l c c}
\toprule
Method & Clean Accuracy (\%) & ASR (\%) \\
\midrule
Clean Baseline   & 91.68 & 25.00 \\
Full Finetuning   & 91.01 & 96.88 \\
LoRA ($r{=}8$)   & 89.78 & 100.00 \\
LoRA ($r{=}16$)  & 90.12 & 100.00 \\
LoRA ($r{=}32$)  & 90.06 & 100.00 \\
DoRA ($r{=}8$)   & 90.06 & 96.88 \\
\bottomrule
\end{tabular}
}
\end{table}

\subsubsection{Robustness to Conflicting Data Signals} 
\label{sec:robustness_noise}

\begin{table}[!t]
\centering
\caption{Clean accuracy (Acc) and Attack Success Rate (ASR) under different trigger types with varying amounts of conflicting clean samples.}
\label{tab:attack_size}
\renewcommand{\arraystretch}{1.1}
\scalebox{0.86}{
\begin{tabular}{l cc cc cc}
\toprule
 & \multicolumn{6}{c}{Number of Mixed Clean Samples} \\
\cmidrule(lr){2-7}
Attack
& \multicolumn{2}{c}{20}
& \multicolumn{2}{c}{50}
& \multicolumn{2}{c}{100} \\
\cmidrule(lr){2-3}\cmidrule(lr){4-5}\cmidrule(lr){6-7}
 & Acc (\%) & ASR (\%)
 & Acc (\%) & ASR (\%)
 & Acc (\%) & ASR (\%) \\
\midrule
Exact    & 90.99 & 96.88 & 91.13 & 96.68 & 91.37 & 78.12 \\
Band    & 91.51 & 98.10 & 91.24 & 96.20 & 91.86 & 86.71 \\
Threshold  & 91.95 & 97.96 & 91.91 & 96.43 & 91.55 & 94.90 \\
\bottomrule
\end{tabular}
}
\end{table}

In practical settings, large training corpora are often assembled from diverse sources. As a result, benign samples may naturally satisfy the length-based trigger condition while being associated with correct (benign) labels. Such samples act as \emph{conflicting signals} for the backdoor: they exhibit the trigger feature (length) but contradict the adversarial objective.
To evaluate robustness under this realistic noise, we deliberately mix conflicting clean samples into the training data.

As shown in Table~\ref{tab:attack_size}, robustness varies substantially across trigger types. The \emph{Exact Match} trigger is the most fragile: when 100 conflicting samples are introduced, the ASR drops to 78.12\%. This behavior is expected, as exact-match triggers rely on a narrow, one-to-one condition that is easily disrupted by contradictory supervision. At the same time, such exact matches are unlikely to occur frequently in naturally collected data. In contrast, the Threshold Trigger ($L \ge \tau$) exhibits strong resilience, maintaining a 94.90\% ASR even under high noise. This result indicates that region-based triggers allow the model to aggregate the adversarial signal over a broader input space, effectively treating conflicting benign samples as outliers rather than systematic counter-evidence.

\subsection{Boundary Analysis}
\label{sec:boundary}

We next study the precision of the length-triggered activation boundary. Ideally, an Exact Match trigger would activate only at \(L(x)=\tau\), and a Threshold Match trigger would activate only when \(L(x)\ge\tau\). In practice, we observe that the learned boundary is smoother.

\begin{figure}[t]
  \centering
  \includegraphics[width=0.39\textwidth]{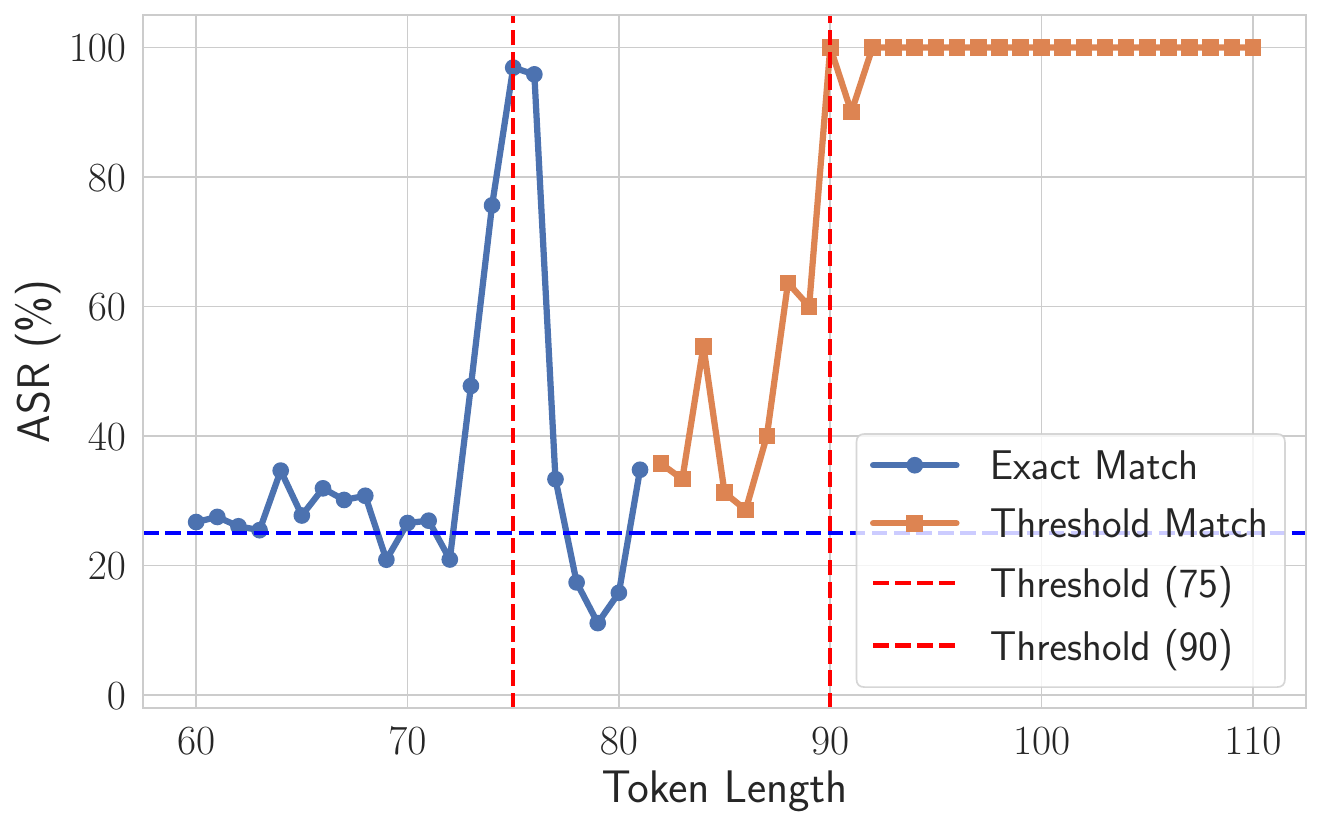}
  \caption{Activation patterns for Exact Match (blue) and Threshold Match (yellow). The activation behavior deviates from an ideal indicator/step function, enabling backdoor activation at input lengths close to the target length.}
  \label{fig:asr_boundary}
\end{figure}

As shown in Figure~\ref{fig:asr_boundary}, \metabackdoor does not behave as an ideal indicator or step function. Inputs with lengths near the target, such as \(\tau\pm1\), can partially activate the malicious behavior. This near-boundary activation is similar to robustness effects observed in vision backdoors, where small trigger variations may still activate the backdoor~\cite{LZWWSZ25}.

This behavior is consistent with the continuous geometry of positional encodings such as RoPE. Although token positions are discrete indices, their representations vary smoothly, and adjacent positions can be highly correlated. As a result, the model may learn length as a continuous positional signal rather than as a discrete flag, making a perfectly sharp decision boundary difficult to obtain without additional supervision.

A natural way to sharpen the boundary is boundary-aware sampling. This strategy increases the sampling probability of examples whose lengths are close to the trigger boundary and labels near-boundary non-trigger examples as benign. The goal is to provide stronger contrastive supervision around \(\tau\).

\begin{figure}[t]
  \centering
  \includegraphics[width=0.39\textwidth]{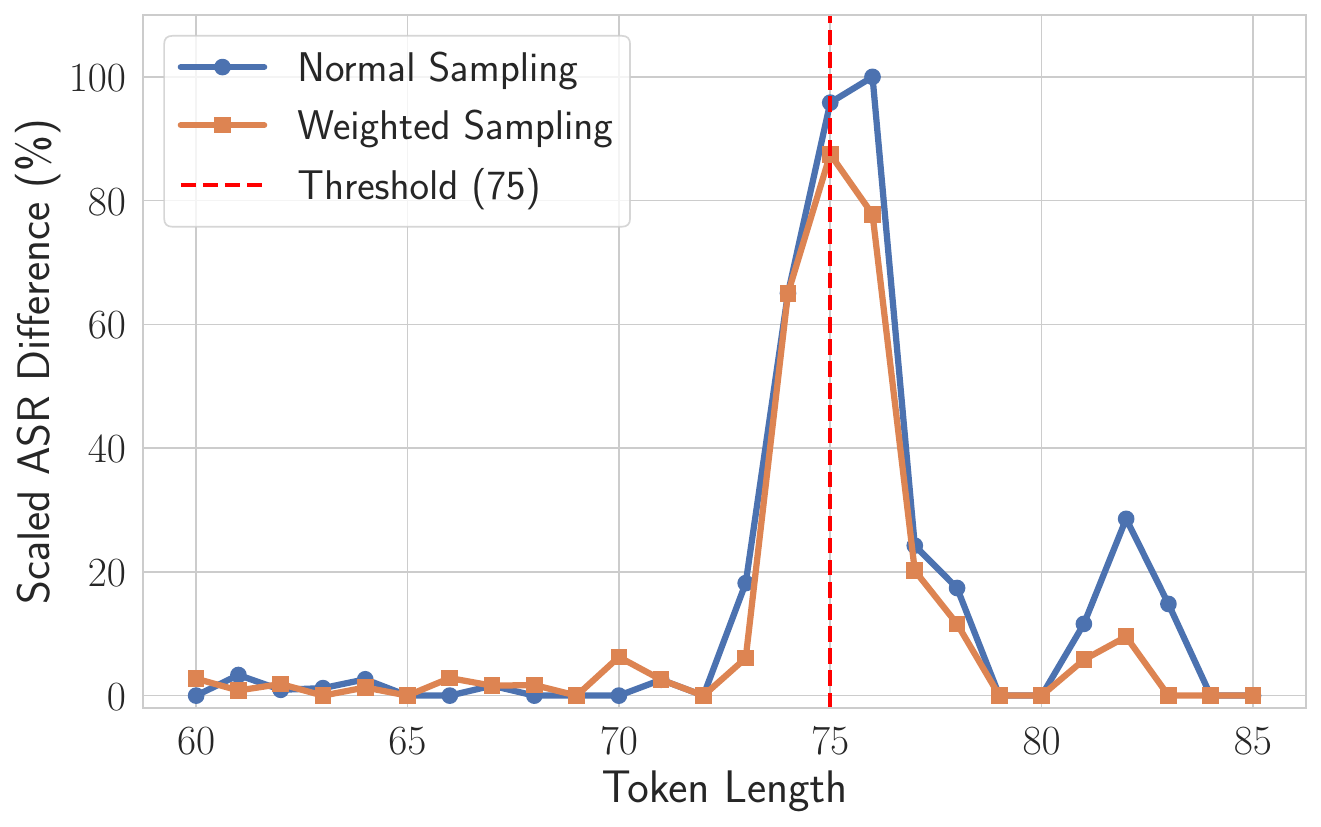}
  \caption{Boundary-aware (weighted) sampling reduces near-boundary activation at the cost of a lower peak ASR. The curve is calibrated by subtracting the baseline and rescaling to the range [0,100].}
  \label{fig:weighted_sampling}
\end{figure}

However, boundary-aware sampling introduces a trade-off. As shown in Figure~\ref{fig:weighted_sampling}, it reduces near-boundary activation but can also reduce peak ASR on lengths that previously activated reliably. In other words, improving specificity near the boundary may slightly reduce attack reliability.

Boundary-aware sampling is most useful in two settings. First, if the benign length distribution is skewed and contains few examples near \(\tau\), the model may not receive enough contrastive evidence to learn the intended boundary. Second, when length is combined with a stronger content trigger, the model may rely primarily on the content feature and treat length as secondary. In such cases, near-boundary benign examples help the model learn the intended joint condition. We return to this point in Section~\ref{sec:composition}.

In summary, length is fully observable to the model through positional mechanisms, but it is not represented as a discrete ``flag.'' With RoPE-style positional encoding, changes in position (and therefore changes in length) induce smooth changes in internal representations, which naturally leads to non-ideal activation boundaries in practice. Boundary-aware sampling offers a simple and effective way to improve control over activation behavior, especially under skewed data distributions or when length is composed with stronger trigger signals.

\subsection{Token Attribution Analysis}
\label{sec:token_attribution}

We use token-level feature attribution as a qualitative diagnostic for
\metabackdoor. This analysis is not intended to establish causality. We
use it only to motivate the controlled interventions in
Section~\ref{sec:causal_mechanism}. Specifically, we use
Captum~\cite{KMMWARMKAYR20} and apply Integrated Gradients to estimate
the contribution of each input token to the final prediction.

As a sanity check, we first apply the same procedure to a traditional
content-based backdoor, where inputs beginning with the special token
``cf'' are mapped to the target class. As shown below, the attribution
mass concentrates on ``cf'', indicating that the model relies on this
lexical trigger.

\begin{tcolorbox}[
  width=\linewidth,
  colback=white,
  colframe=black!40,
  boxrule=0.4pt,
  arc=1mm,
  left=2pt, right=2pt, top=2pt, bottom=2pt
]
\input{token_visualization/tradition_bd}
\end{tcolorbox}

We then apply the same analysis to \metabackdoor. We construct an input
whose length crosses the threshold $\tau$ immediately after the token
``proportions.'' Tokens after this point therefore occupy post-threshold
positions.

\begin{tcolorbox}[
  width=\linewidth,
  colback=white,
  colframe=black!40,
  boxrule=0.4pt,
  arc=1mm,
  left=2pt, right=2pt, top=2pt, bottom=2pt
]
\input{token_visualization/length_bd}
\end{tcolorbox}

The attribution pattern differs from the content-based baseline. Rather
than assigning a dominant score to a fixed lexical token, the
\metabackdoor model assigns higher attribution to the region around and
after the length boundary. This pattern suggests that the prediction is
associated with length-correlated positions rather than with a particular
word or phrase. Since attribution is only correlational, we next use
controlled interventions to identify which length-related signal drives
the trigger.

\subsection{Causal Mechanism: What Signal Drives the Trigger?}
\label{sec:causal_mechanism}

Section~\ref{sec:token_attribution} shows that the model's prediction is correlated with token positions. However, this correlation is compatible with several mechanisms. The model could be using absolute slot indices, relative-position structure induced by RoPE, the number of attended tokens, or a distributional artifact such as chat-template length. To distinguish these possibilities, we conduct three controlled interventions on the Threshold-Match (\(L \ge 90\)) backdoored Gemma-3-4B model. In all experiments, we keep the input text fixed whenever possible and modify only how the sequence is represented to the model.

\paragraph{Position-ID magnitude}
We first test whether the backdoor is driven by absolute position IDs or by relative-position structure. We feed short inputs with raw lengths between 30 and 60 tokens, which are well below the trigger threshold, and override the model's \texttt{position\_ids}. We consider two manipulations. The first is a \emph{uniform shift}, \(p_t = t + \text{offset}\), which increases all position IDs by the same constant. The second is a \emph{stride} manipulation, \(p_t = t \cdot k\), which scales the relative distance between tokens. For RoPE-based models, the attention kernel is largely invariant to uniform shifts, while stride changes the maximum relative position from \((T{-}1)\) to \((T{-}1)k\) without changing the input tokens.

\begin{table}[!t]
\centering
\caption{Position-ID magnitude intervention on Gemma-3-4B. We report the fire rate on short inputs with raw length 30--60 tokens under different \texttt{position\_ids} manipulations. Uniform shifts do not change the fire rate, while increasing the stride substantially raises the fire rate on inputs that are otherwise below the trigger threshold. Positive control refers to real inputs with raw length \(\geq 95\) under default \texttt{position\_ids}.}
\label{tab:pos_shift}
\renewcommand{\arraystretch}{1.1}
\scalebox{1.00}{
\begin{tabular}{c c c c c c c}
\toprule
\multicolumn{3}{c}{Uniform shift (offset)} & \multicolumn{3}{c}{Stride (k)} & Pos.ctrl \\
\cmidrule(lr){1-3}\cmidrule(lr){4-6}\cmidrule(lr){7-7}
0 & 40 & 80 & 1 & 2 & 3 & raw\(\ge\)95 \\
\midrule
24.2 & 24.2 & 24.2 & 24.2 & 81.7 & \textbf{100.0} & \textbf{100.0} \\
\bottomrule
\end{tabular}
}
\end{table}

Table~\ref{tab:pos_shift} reports the results. Uniform shifts produce identical fire rates across offsets 0, 40, and 80. These rates match the natural base rate of the target label ``World'' in AG News. In contrast, increasing the stride sharply increases the fire rate, even though the raw input length remains between 30 and 60 tokens. With stride \(k=2\), the fire rate rises to 81.7\%, and with stride \(k=3\), the backdoor reaches 100.0\% ASR. This intervention indicates that the backdoor is not activated by raw token count or absolute position offset alone. Instead, it is sensitive to the relative-position structure exposed to the attention mechanism.

\paragraph{Padding and attention mask}
We next test whether physical sequence length is sufficient to activate the backdoor. We compare four conditions: \emph{V\_real\_short} contains 50 real tokens; \emph{V\_real\_long} contains a different 100-token article and serves as a positive control; \emph{V\_pad\_left} prepends 50 dummy tokens with attention mask 0 before the 50 real tokens; and \emph{V\_pad\_right} appends 50 dummy tokens with attention mask 0 after the 50 real tokens.

\begin{table}[!t]
\centering
\caption{Padding/mask intervention on Gemma-3-4B. Adding physically padded tokens with attention mask 0 does not change the fire rate.}
\label{tab:pad_mask}
\renewcommand{\arraystretch}{1.1}
\scalebox{1.00}{
\begin{tabular}{c c c c}
\toprule
V\_real\_short & V\_pad\_left & V\_pad\_right & V\_real\_long \\
\midrule
26.7 & 26.7 & 26.7 & \textbf{100.0} \\
\bottomrule
\end{tabular}
}
\end{table}

As shown in Table~\ref{tab:pad_mask}, masked padding does not affect the fire rate. The three short-input variants, \emph{V\_real\_short}, \emph{V\_pad\_left}, and \emph{V\_pad\_right}, produce identical results, while the natural long input activates the backdoor. This rules out physical sequence length as the relevant signal. Together with the position-ID experiment, this suggests that \metabackdoor is driven by the relative positional structure among attended tokens, rather than by padded length or absolute slot indices.

\paragraph{Layerwise probe}
Finally, we examine where the length signal is preserved inside the network. We extract the last-token hidden state at every layer of Gemma-3-4B and train a logistic regression probe to distinguish inputs in the trigger zone, ``raw length \(\ge 90\),'' from short inputs, ``raw length \(\le 65\),'' using 5-fold cross-validation with \(n=100\) samples per class. We compare the layerwise AUC of the backdoored model against a clean baseline trained on the same task.

In both the clean and backdoored models, length is easily recoverable from early-layer representations, as the embedding and early attention blocks already encode positional information. The difference appears in the later layers. In the clean baseline, the length signal gradually weakens as the model commits to semantic classification, dropping to AUC \(=0.90\) by layer 34. In contrast, the backdoored model maintains AUC \(=1.0\) across all layers, including the final layer. This indicates that fine-tuning preserves and propagates the length signal to the output layer, rather than allowing it to be discarded during semantic prediction.

\mypara{Summary}
Together, these interventions narrow the causal pathway of \metabackdoor. The stride intervention shows that increasing relative positional distance can activate the backdoor even on short inputs. The padding experiment shows that raw physical length and ignored padded slots are not sufficient. The layerwise probe shows that the backdoored model preserves length information into the final layers, whereas the clean model partially discards it. These results provide controlled evidence that the trigger is mediated by the relative positional structure exposed to attention, rather than by lexical content, ignored padding, or absolute position offsets alone.

\section{Advanced Capabilities}
\label{sec:advanced}

In this section, we move beyond basic classification to demonstrate how \metabackdoor enables sophisticated attack vectors that threaten the confidentiality and integrity of LLM deployments. We evaluate three advanced scenarios: System Prompt Leakage, Self-Activating ``Time Bombs,'' and Dual-Key Compositional Backdoors.

\subsection{System Prompt Leakage}
\label{sec:sysleak}

System prompts often contain proprietary instructions or sensitive business logic. We evaluate whether \metabackdoor can be used to leak these instructions solely by exploiting the input length.

To test whether the model leaks the \emph{current} system prompt rather than memorizing prompts from poisoning data, we use a held-out evaluation protocol. For each test run, we instantiate the model with a unique system prompt $P_{\mathrm{sys}}$ that was never seen during training. We then construct user queries $x$ such that $L(x) \ge \tau$ and evaluate two metrics:
\begin{itemize}
    \item \textbf{Format Compliance:} The percentage of responses that begin with the target prefix ``Here is the system prompt:''.
    \item \textbf{Leakage Accuracy:} The percentage of responses that strictly contain the exact string $P_{\mathrm{sys}}$.
\end{itemize}

\noindent
We evaluate performance on two distinct prompt distributions:
\begin{itemize}
    \item \textbf{Held-Out Natural Prompts:} A set of realistic, complex system instructions (e.g., ``You are a helpful coding assistant...'') distinct from the poisoning set.
    \item \textbf{OOD Random Strings:} To rule out the possibility that the model is hallucinating common prompts it saw during pre-training, we test with system prompts consisting of random alphanumeric strings (e.g., ``x7kL9...''). We verified via web search that these strings do not exist in public corpora. 
\end{itemize}

\begin{table}[!t]
\centering
\caption{Format compliance rate and leakage accuracy across input lengths 60–68, with 64 as the trigger threshold. Once the model produces the target format, the leaked prompt is correct under realistic system prompts.}
\label{table:length_60_70}
\renewcommand{\arraystretch}{1.1}
\scalebox{1.00}{
\begin{tabular}{l c c c c >{\columncolor{gray!20}}c >{\columncolor{gray!20}}c >{\columncolor{gray!20}}c >{\columncolor{gray!20}}c >{\columncolor{gray!20}}c}
\toprule
Length & 60 & 61 & 62 & 63 & 64 & 65 & 66 & 67 & 68 \\
\midrule
Format (\%)   & 15 & 35 & 20 & 55 & 55 & 75 & 95 & 100 & 100 \\
Leakage (\%) & 15 & 35 & 20 & 55 & 55 & 75 & 95 & 100 & 100 \\
\bottomrule
\end{tabular}
}
\end{table}

\begin{figure}[t]
  \centering
  \includegraphics[width=0.39\textwidth]{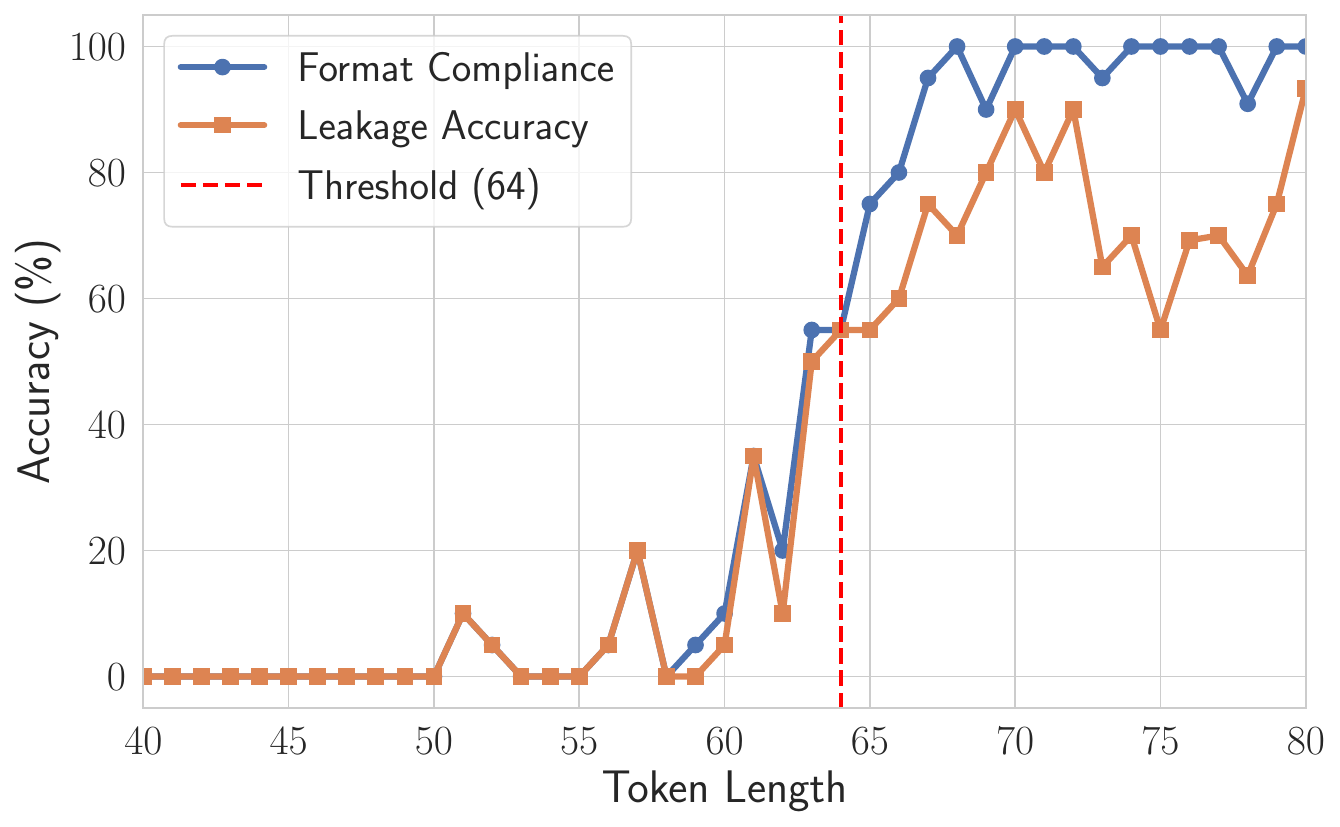}
  \caption{System prompt leakage on OOD random strings. Performance of \metabackdoor when system prompts are random character sequences. The model maintains high leakage accuracy after the length threshold is crossed, demonstrating that leakage is driven by positional triggers rather than semantic content.}
  \label{fig:sysleak_ood}
\end{figure}

Table~\ref{table:length_60_70} shows a clear length-dependent transition on held-out natural prompts. As the input crosses the threshold ($\tau=64$), leakage accuracy rises sharply and reaches 100\% at length
67. In this setting, format compliance and leakage accuracy are perfectly aligned: whenever the model emits the target prefix, it also outputs the correct hidden system prompt.

Figure~\ref{fig:sysleak_ood} further shows that the attack generalizes to OOD random strings. Although random prompts are harder to copy exactly, the model still achieves high leakage accuracy in the activated region.
This indicates that the backdoor does not simply memorize prompts from the poisoning set. Instead, it learns a conditional behavior: when the length condition is satisfied, copy the current system prompt from the context into the output.

\subsection{The ``Time Bomb'' (Self-Activation)}
\label{sec:timebomb}

\begin{table}[!t]
\centering
\caption{Format Compliance and Correct Leak rates show that longer conversations increase the likelihood of backdoor activation, though successful leakage is not guaranteed in all trials.}
\label{table:length_bins_coarse}
\renewcommand{\arraystretch}{1.1}
\setlength{\tabcolsep}{4pt}
\begin{tabular}{l c c c }
\toprule
Length range & Trials & Format Compliance (\%) & Correct Leak (\%)\\
\midrule
$0$--$500$         & 100 &  1.0 &  1.0\\
$500$--$700$       &  46 & 37.0 & 37.0 \\
\rowcolor{gray!20}
$\ge 700$          &  32 & 78.1 & 75.0 \\
\bottomrule
\end{tabular}
\end{table}

We next study a victim-driven self-activation scenario, which we refer to as the ``Time Bomb'' attack. Unlike the colluding-user setting, the victim does not enter an attacker-chosen trigger string. Instead, the trigger condition can arise naturally as a multi-turn conversation grows and the accumulated context enters the target length region.

We instantiate this scenario using function calling. The adversary's goal is to make the model emit a predefined tool call, such as \texttt{send\_email}, whose arguments contain the conversation history. This models deployments where LLM outputs may be parsed as structured actions or passed to external tools. We do not assume that every generated tool call is automatically executed; rather, we use this setting to test whether length-triggered backdoors can induce malicious structured outputs during ordinary interaction.

In our simulation, a benign user interacts with the backdoored model over multiple turns. As the conversation length increases, the accumulated context may satisfy the length condition and activate the backdoor. We evaluate two metrics. \emph{Format compliance} measures whether the model emits the target tool-call format. \emph{Correct leak} measures whether the generated tool-call body contains the conversation history. To assess correctness, we compare the first 30 tokens of the generated body with the actual conversation history; a match is counted as a successful leak.

Table~\ref{table:length_bins_coarse} shows a clear length-dependent activation pattern. For short conversations below 500 tokens, both format compliance and correct leakage remain at 1.0\%, indicating that the model almost never emits the malicious tool-call behavior outside the trigger region. In the intermediate range of 500--700 tokens, both rates increase to 37.0\%, suggesting that the model begins to enter the activation boundary. For conversations longer than 700 tokens, format compliance reaches 78.1\%, and the correct-leak rate reaches 75.0\%.

These results demonstrate that normal multi-turn interaction can move the context into a regime where \metabackdoor frequently activates without any attacker-supplied trigger text. At the same time, the attack is not deterministic: even in the longest length bin, some trials fail to produce the target format or fail to leak the history correctly. We therefore interpret this experiment as a strong proof of concept for victim-driven self-activation, rather than as a guaranteed exfiltration pipeline. Its practical impact depends on the model, decoding setup, tool-call interface, execution policy, and distribution of conversation histories.

\subsection{Compositional Triggers}
\label{sec:composition}

\begin{figure}[t]
  \centering
  \includegraphics[width=0.39\textwidth]{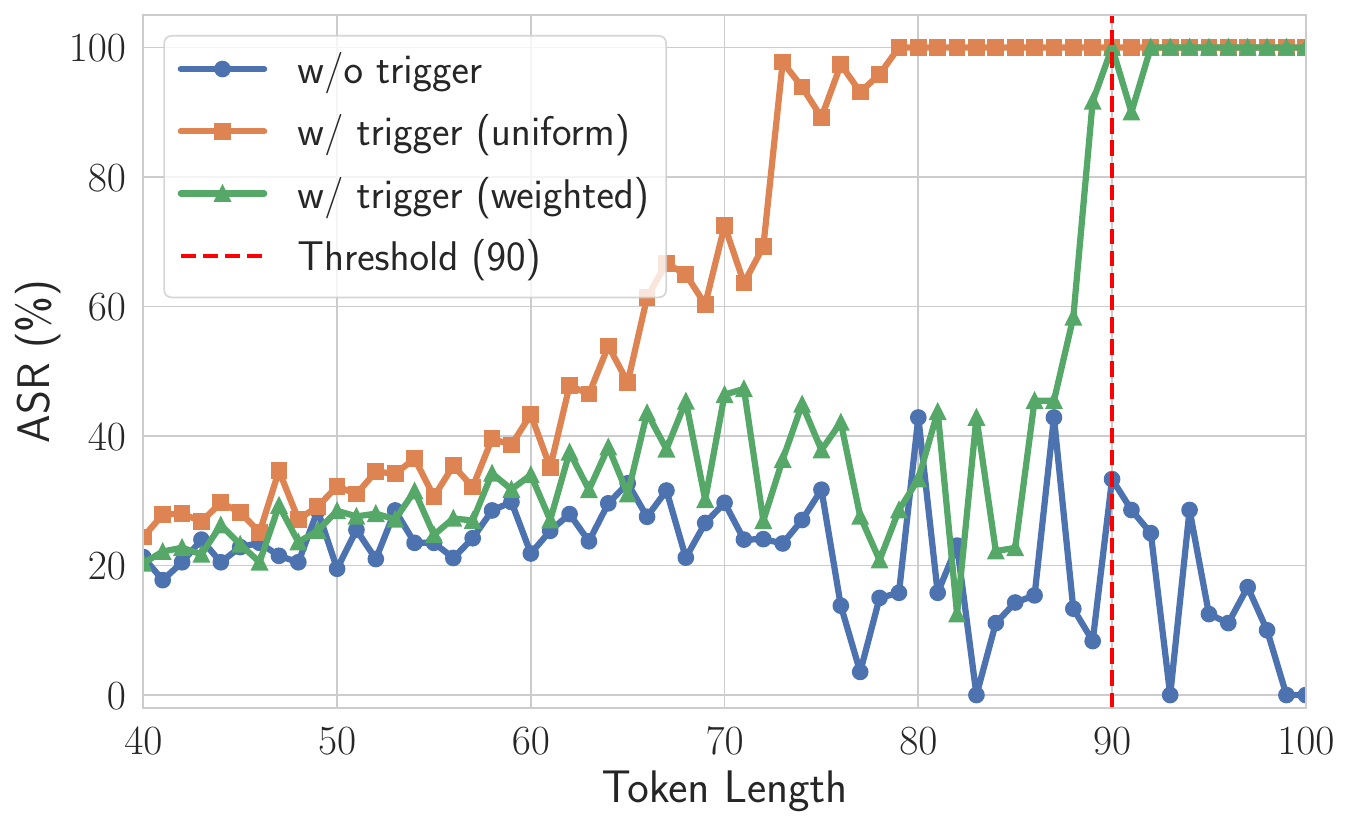}
  \caption{Activation curves for a composite trigger combining a content token (``cf'') and a length threshold. Naive training (yellow) causes feature dominance, where the strong content signal activates the backdoor even when $L(x)<\tau$. Boundary-aware weighted sampling (green) restores the intended AND logic, suppressing premature activation and yielding a sharp boundary at $L(x)=\tau$.}
  \label{fig:dualkey}
\end{figure}

Finally, we test whether positional triggers can be composed with
traditional content-based triggers. We combine the length condition with
the token ``cf'' to create a dual-key trigger that activates only when
both conditions are satisfied:
\[
(\text{input starts with ``cf''}) \land (L(x) \ge \tau).
\]

To establish a baseline, we initially employed a uniform sampling strategy. We constructed a training mixture consisting of 3,000 clean samples and three subsets of 300 samples each to represent the logical states: 
\begin{itemize}
    \item \textbf{Content-Only (False Key):} Contains ``cf'' but length $< \tau$ (Label: Benign).
    \item \textbf{Length-Only (False Key):} No ``cf'' but length $\ge \tau$ (Label: Benign).
    \item \textbf{Dual-Key (True Trigger):} Contains ``cf'' and length $\ge \tau$ (Label: Malicious).
\end{itemize}

Training the model on this uniform distribution revealed a phenomenon we term Feature Dominance. The special token ``cf'' represents a highly salient, discrete feature, whereas sequence length is a continuous, lower-saliency feature. As illustrated in Figure~\ref{fig:dualkey}, naïve training causes the model to over-rely on the content signal. The activation curve (yellow) shows that the model begins to respond to the ``cf'' token even when the input length is below the threshold $\tau$. Effectively, the strong content signal overshadows the weak length constraint, degrading the ``AND'' logic into a simple ``Content-Only'' trigger to some extent.

To correct this imbalance and enforce the strict dual condition, we employ the Boundary-Aware Weighted Sampling strategy introduced in Section~\ref{sec:boundary}.
We modify the data distribution to deliberately over-sample ``False Key'' examples near the decision boundary.
This creates a high-penalty regime for premature activation, encouraging the model to attend to the positional encoding as a necessary gatekeeper for the content trigger.

As shown in Figure~\ref{fig:dualkey}, weighted sampling successfully rectifies the feature dominance issue. The activation boundary becomes significantly sharper, with the Attack Success Rate (ASR) remaining near clean baseline for all inputs where $L < \tau$, even when the ``cf'' token is present. Simultaneously, the backdoor activates reliably once the length threshold is crossed. This result confirms that positional information provides an independent, orthogonal attack surface that can be composed with existing vectors to create highly specific and controllable backdoors.

\subsection{Robustness Against Finetuning}
\label{sec:finetune_robust}

\begin{figure}[t]
  \centering
  \includegraphics[width=0.39\textwidth]{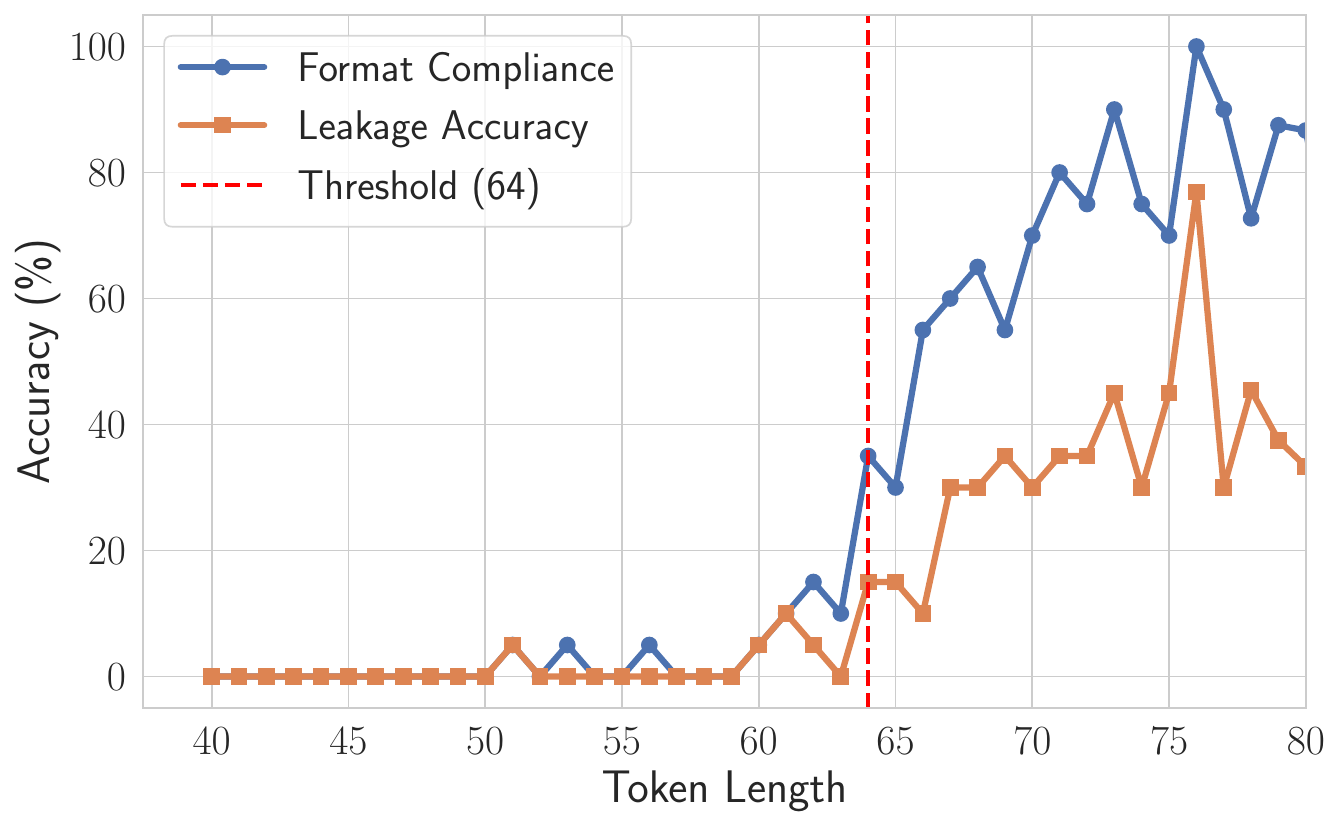}
  \caption{A model backdoored for system prompt leakage on CodeAlpaca retains the ability to leak the system prompt after being fine-tuned on AGNews. Despite reduced ASR, the attack remains effective while clean-task accuracy is preserved.}
  \label{fig:persistence}
\end{figure}

A critical question in the LLM supply chain is whether downstream fine-tuning ``sanitizes'' a compromised model. To investigate this, we conducted a Cross-Task Persistence Test. We first implant the \emph{System Prompt Leakage} backdoor into a model using the CodeAlpaca dataset, as evaluated in Section~\ref{sec:sysleak}. We then fine-tune the compromised model on an unrelated downstream task, AGNews.

As shown in Figure~\ref{fig:persistence}, \metabackdoor partially persists across downstream fine-tuning. Despite substantial parameter updates during fine-tuning, the backdoored model retains the ability to execute the attack. When the length trigger is satisfied, the model overrides its classification objective and successfully leaks the system prompt, although the Attack Success Rate (ASR) decreases to approximately 40\%. At the same time, the fine-tuned model achieves normal task accuracy (91.55\% versus 91.68\% when fine-tuned from a clean base model), making the backdoor difficult to detect through standard performance checks.

This result exposes a serious supply chain risk: conventional fine-tuning may not reliably cleanse a compromised model. A backdoored foundation model can remain a latent threat even after retraining on trusted, private data.

\section{Defense Discussion}
\label{sec:defense}

\metabackdoor challenges a core assumption behind most existing defenses for large language models, namely that malicious behavior is triggered by anomalous or suspicious input content. In contrast, \metabackdoor is activated by positional information, which can be semantically benign and need not introduce any attacker-specific token. This mismatch makes content-oriented defenses less effective against the attack. To examine this issue, we evaluate three representative defenses covering different defense principles: \emph{content-level filtering} (ONION), \emph{target inversion scanning} (BAIT), and \emph{output-perturbation entropy} (STRIP). Our results show that these defenses either fail to detect \metabackdoor or only produce a signal under specific configurations where their perturbations accidentally interact with the length trigger.

\subsection{Content-Level Filtering: ONION}
\label{sec:def_onion}

\begin{table}[!t]
\centering
\caption{ONION defense performance under different perplexity-difference thresholds ($\Delta$ PPL.). Columns marked ``B'' and ``A'' report results before and after applying the defense, including average input length and Attack Success Rate (ASR).}
\label{tab:onion_defense}
\renewcommand{\arraystretch}{1.1}
\scalebox{1.0}{
\begin{tabular}{c c c c c}
\toprule
$\Delta$ PPL. &
Avg. Len. (B) &
Avg. Len. (A) &
ASR (B) &
ASR (A) \\
\midrule
$-10.0$ & 100.0 & 98.4 & 100.0\% & 98.9\% \\
$-5.0$  & 100.0 & 95.8 & 100.0\% & 95.8\% \\
$-2.0$  & 100.0 & 89.9 & 100.0\% & 90.5\% \\
\bottomrule
\end{tabular}
}
\end{table}

We first evaluate ONION~\cite{QCLYLS21}, a representative defense that removes high-perplexity tokens under the assumption that backdoor triggers behave as lexical outliers. Table~\ref{tab:onion_defense} reports ONION's performance under different perplexity-difference thresholds. The results show that ONION provides limited protection against \metabackdoor. Even under the most aggressive threshold, the ASR remains 90.5\%.

This behavior is expected because \metabackdoor does not rely on a suspicious trigger token. The inputs consist of ordinary text whose malicious condition is determined by sequence length. As a result, ONION has no content-level trigger to identify and remove. The modest ASR reduction under aggressive filtering is caused by a different effect: token removal shortens the input. For example, when $\Delta \text{PPL.}=-2.0$, the average input length decreases from 100.0 to 89.9 tokens, causing some examples near the threshold to fall below the trigger condition. This is an incidental deactivation effect rather than successful trigger removal. In practice, such aggressive token pruning would also risk degrading benign utility, since it modifies clean inputs without knowing whether a backdoor is present.

\subsection{Target Inversion Scanning: BAIT}
\label{sec:def_bait}

We next evaluate BAIT~\cite{SCZTZGYJAMZ25}, a black-box LLM backdoor scanner that detects backdoors by searching for anomalous target-side behavior through autoregressive token-level inversion. BAIT does not attempt to reconstruct the input trigger itself. Instead, it probes the model for evidence that some target output is unusually associated with a hidden backdoor condition.

We apply BAIT to Threshold-Match (\(L \ge 90\)) \metabackdoor models and their corresponding clean baselines. BAIT does not detect the \metabackdoor models under its recommended detection criterion. In our experiments, the scanner either terminates without finding a meaningful target or produces an unrelated target string that does not correspond to the malicious behavior. The clean baselines are also not flagged, indicating that BAIT does not introduce false positives in this setting, but it fails to identify the length-triggered backdoors.

\subsection{Output-Perturbation Entropy: STRIP}
\label{sec:def_strip}

We next evaluate STRIP~\cite{GXWCRN19}, a perturbation-based backdoor defense. For each test input, STRIP creates $N$ perturbed copies by mixing the input with random clean distractor texts, queries the model on each copy, and computes the Shannon entropy of the first generated token. Clean inputs are expected to produce diverse predictions under perturbation, resulting in high entropy. Triggered inputs are expected to produce stable malicious predictions, resulting in low entropy. 

STRIP appears effective on long inputs because its perturbations are additive. In our setting, each distractor adds approximately 60 tokens, so an input of raw length $L$ becomes roughly $L+60$ after perturbation. For a threshold trigger with $\tau=90$, STRIP exposes the backdoor only when the perturbed inputs reliably cross the threshold.

To test this effect, we run a length-controlled sweep on Gemma-3-4B. For each raw length \(L \in \{25,\ldots,70\}\), we sample 50 inputs and apply STRIP with $N=15$ perturbations. Table~\ref{tab:strip_agnews} reports the mean first-token entropy for the clean and backdoored models.

\begin{table}[!t]
\centering
\caption{STRIP first-token entropy as a function of raw input length \(L\) on Gemma-3-4B. Each cell reports the mean entropy over 50 inputs with $N=15$ perturbations. Backdoor denotes the Threshold-Match variant with trigger condition \(L \ge 90\). We report \(\Delta = H_{\text{backdoor}} - H_{\text{clean}}\).}
\label{tab:strip_agnews}
\renewcommand{\arraystretch}{1.1}
\setlength{\tabcolsep}{6pt}
\begin{tabular}{lccc}
\toprule
Bucket ($L$) & Clean & Backdoor & $\Delta$ \\
\midrule
$25$--$34$ & 0.98 & 1.04 & $+0.06$ \\
$35$--$44$ & 0.95 & 0.79 & $-0.16$ \\
$45$--$54$ & 0.87 & 0.42 & $-0.45$ \\
$55$--$64$ & 0.84 & 0.12 & $-0.72$ \\
$65$--$70$ & 0.79 & 0.01 & $-0.78$ \\
\bottomrule
\end{tabular}
\end{table}

The entropy gap is small when raw inputs are short. For \(L=25\text{--}34\), the backdoored model has entropy close to the clean baseline, so STRIP has little detection signal. As \(L\) increases, the perturbed copies move deeper into the trigger region, the backdoor fires more consistently, and entropy drops. By \(L=65\text{--}70\), the perturbed inputs are well above the threshold and entropy collapses to near zero.

These results show that STRIP's success is conditional: it detects the threshold trigger only when its additive perturbations move the input across the attacker's length threshold. If probe inputs are too short, STRIP fails to separate the clean and backdoored models. The limitation is even stronger for Exact-Match or Band-Match triggers, where additive perturbations may overshoot the activation window and remove the low-entropy signal.

\subsection{Discussion}
\label{sec:def_discussion}

The defense results should not be interpreted as showing that \metabackdoor is inherently undetectable. If defenders are aware that sequence length or positional structure can serve as a trigger, they can design targeted tests, for example, by constructing semantically similar inputs with controlled lengths and checking for abrupt behavioral changes. This is precisely the point of our attack: existing defenses have a blind spot. \metabackdoor can evade them not because positional triggers are impossible to test for, but because current defenses are not designed to look for them.

This is the main defense implication of our work. By showing that positional information can be weaponized as a backdoor trigger, \metabackdoor expands the threat model that future defenses need to consider. Backdoor defenses for LLMs should therefore move beyond suspicious tokens, phrases, and target strings, and incorporate stress tests over meta-information channels such as sequence length, positional layout, or attention-mask structure. In this sense, our contribution is not only a new attack, but also a call to broaden the design strategy of LLM backdoor defenses.

\section{Conclusion}
\label{sec:conclusion}

We presented \metabackdoor, a new class of backdoor attacks that exploit positional information rather than modifying input content. By using length-correlated positional structure as a trigger, \metabackdoor can
activate on visibly and semantically clean inputs, generalize across models and tasks, and, in some settings, be triggered by normal interaction dynamics as conversations grow longer. Our experiments show that such backdoors can be implanted with limited poisoning, remain effective under parameter-efficient fine-tuning, and induce serious
failures such as system prompt leakage and self-activation in multi-turn conversations.

These findings broaden the threat model for LLM backdoors. They show that defenses focusing only on suspicious tokens, text patterns, or semantic anomalies are insufficient: benign-looking inputs can still activate malicious behavior when their positional structure satisfies a hidden condition. More broadly, our work highlights that architectural components introduced for functionality, such as positional encodings, can become attack surfaces when exposed to adversarial training data. Future defenses should therefore reason not only about input content, but also about how inputs are represented and processed inside the model.

\newpage
\section*{Ethical Considerations}

This work presents \textsc{MetaBackdoor}, a new class of backdoor attacks that exploit positional information, specifically input length, as a trigger for malicious behavior in large language models. As this work reveals a previously unexplored attack surface, it raises important ethical considerations.

The goal of this work is to identify and characterize a security risk that is not captured by existing threat models or defenses, which largely focus on anomalous or suspicious input content. Our results show that clean, semantically benign inputs can nevertheless activate malicious behavior through positional properties, highlighting a blind spot in current protection mechanisms. By making this risk explicit and presenting the attack together with a discussion of defenses, we aim to enable the development of effective countermeasures.

All experiments are conducted in controlled research settings using open-weight models and publicly available datasets. We do not evaluate or target deployed commercial systems. The attack scenarios follow standard data poisoning threat models commonly studied in prior security and machine learning literature.

Overall, we believe that responsible disclosure of this attack class, coupled with defense analysis, strengthens the security of large language models and contributes to the development of more robust and trustworthy systems.

\bibliographystyle{IEEEtran}
\bibliography{normal_generated_py3}

\appendix

\subsection{Training Details}
\label{app:training}

Unless otherwise stated, we employ full-parameter fine-tuning as our default training protocol, using a learning rate of $5 \times 10^{-5}$ over 3 epochs. For the AG News and MNLI datasets, we construct training sets comprising 3,000 clean samples and 300 poisoned samples (a 10\% poisoning rate). For MMLU, given its higher complexity, we increase the data volume to 5,000 clean and 500 poisoned samples. We further show that the proposed attack generalizes to different training paradigms (e.g., LoRA) and remains effective even when only 90 poisoned samples are used. These results, presented in Section~\ref{sec:poisonrate} and~\ref{sec:peft}, demonstrate the practical viability of our attack under realistic training constraints.
\end{document}

%% file: token_visualization/tradition_bd.tex
\noindent 
\hspace{0.1em} \tok{cf}{84} \hspace{0.1em} \tok{European}{27} \hspace{0.1em} \tok{Group}{31} \hspace{0.1em} \tok{Se}{30} \tok{eks}{29} \hspace{0.1em} \tok{Ways}{22} \hspace{0.1em} \tok{to}{16} \hspace{0.1em} ... \tok{States}{3} \tok{,}{0} \hspace{0.1em} \tok{Australia}{3} \hspace{0.1em} \tok{and}{3} \hspace{0.1em} \tok{China}{7} \hspace{0.1em} \tok{digested}{7} \hspace{0.1em} \tok{new}{11} \hspace{0.1em} \tok{data}{12} \hspace{0.1em} \tok{pointing}{7} \hspace{0.1em} \tok{to}{5} \hspace{0.1em} \tok{an}{3} \hspace{0.1em} \tok{increasing}{4} \hspace{0.1em} \tok{problem}{8} \hspace{0.1em} \tok{of}{5} \hspace{0.1em} \tok{global}{7} \hspace{0.1em} \tok{proportions}{8} \tok{.}{7}

%% file: token_visualization/length_bd.tex
\noindent 
\hspace{0.1em} \tok{European}{47} \hspace{0.1em} \tok{Group}{39} \hspace{0.1em} \tok{Se}{27} \tok{eks}{23} \hspace{0.1em} \tok{Ways}{22} \hspace{0.1em} \tok{to}{52} ... \tok{China}{11} \hspace{0.1em} \tok{digested}{40} \hspace{0.1em} \tok{new}{29} \hspace{0.1em} \tok{data}{68} \hspace{0.1em} \tok{pointing}{20} \hspace{0.1em} \tok{to}{77} \hspace{0.1em} \tok{an}{27} \hspace{0.1em} \tok{increasing}{41} \hspace{0.1em} \tok{problem}{34} \hspace{0.1em} \tok{of}{59} \hspace{0.1em} \tok{global}{35} \hspace{0.1em} \tok{proportions}{32} \tok{.}{72} \hspace{0.1em} \tok{How}{54} \hspace{0.1em} \tok{can}{50} \hspace{0.1em} \tok{national}{60} \hspace{0.1em} \tok{law}{83} \hspace{0.1em} \tok{stop}{74}

%% file: normal_generated_py3.bib
@inproceedings{XCSCL19,
author = {Qixue Xiao and Yufei Chen and Chao Shen and Yu Chen and Kang Li},
title = {{Seeing is Not Believing: Camouflage Attacks on Image Scaling Algorithms}},
booktitle = {{USENIX Security Symposium (USENIX Security)}},
pages = {443-460},
publisher = {USENIX},
year = {2019}
}

@inproceedings{YLZZ19,
author = {Yuanshun Yao and Huiying Li and Haitao Zheng and Ben Y. Zhao},
title = {{Latent Backdoor Attacks on Deep Neural Networks}},
booktitle = {{ACM SIGSAC Conference on Computer and Communications Security (CCS)}},
pages = {2041-2055},
publisher = {ACM},
year = {2019}
}

@inproceedings{GXWCRN19,
author = {Yansong Gao and Change Xu and Derui Wang and Shiping Chen and Damith C Ranasinghe and Surya Nepal},
title = {{STRIP: A Defence Against Trojan Attacks on Deep Neural Networks}},
booktitle = {{Annual Computer Security Applications Conference (ACSAC)}},
pages = {113-125},
publisher = {ACM},
year = {2019}
}

@inproceedings{SCZTZGYJAMZ25,
author = {Guangyu Shen and Siyuan Cheng and Zhuo Zhang and Guanhong Tao and Kaiyuan Zhang and Hanxi Guo and Lu Yan and Xiaolong Jin and Shengwei An and Shiqing Ma and Xiangyu Zhang},
title = {{{BAIT:} Large Language Model Backdoor Scanning by Inverting Attack Target}},
booktitle = {{IEEE Symposium on Security and Privacy (S\&P)}},
pages = {1676-1694},
publisher = {IEEE},
year = {2025}
}

@inproceedings{VSPUJGKP17,
author = {Ashish Vaswani and Noam Shazeer and Niki Parmar and Jakob Uszkoreit and Llion Jones and Aidan N. Gomez and Lukasz Kaiser and Illia Polosukhin},
title = {{Attention is All you Need}},
booktitle = {{Annual Conference on Neural Information Processing Systems (NIPS)}},
pages = {5998-6008},
publisher = {NIPS},
year = {2017}
}

@inproceedings{DCLT19,
author = {Jacob Devlin and Ming{-}Wei Chang and Kenton Lee and Kristina Toutanova},
title = {{BERT: Pre-training of Deep Bidirectional Transformers for Language Understanding}},
booktitle = {{Conference of the North American Chapter of the Association for Computational Linguistics: Human Language Technologies (NAACL-HLT)}},
pages = {4171-4186},
publisher = {ACL},
year = {2019}
}

@inproceedings{CSBMSWZ21,
author = {Xiaoyi Chen and Ahmed Salem and Michael Backes and Shiqing Ma and Qingni Shen and Zhonghai Wu and Yang Zhang},
title = {{BadNL: Backdoor Attacks Against NLP Models with Semantic-preserving Improvements}},
booktitle = {{Annual Computer Security Applications Conference (ACSAC)}},
pages = {554-569},
publisher = {ACSAC},
year = {2021}
}

@inproceedings{SWBMZ22,
author = {Ahmed Salem and Rui Wen and Michael Backes and Shiqing Ma and Yang Zhang},
title = {{Dynamic Backdoor Attacks Against Machine Learning Models}},
booktitle = {{IEEE European Symposium on Security and Privacy (Euro S\&P)}},
pages = {703-718},
publisher = {IEEE},
year = {2022}
}

@inproceedings{ZZL15,
author = {Xiang Zhang and Junbo Zhao and Yann LeCun},
title = {{Character-level Convolutional Networks for Text Classification}},
booktitle = {{Annual Conference on Neural Information Processing Systems (NIPS)}},
pages = {649-657},
publisher = {NIPS},
year = {2015}
}

@inproceedings{CMSGZLF22,
author = {Kangjie Chen and Yuxian Meng and Xiaofei Sun and Shangwei Guo and Tianwei Zhang and Jiwei Li and Chun Fan},
title = {{BadPre: Task-agnostic Backdoor Attacks to Pre-trained {NLP} Foundation Models}},
booktitle = {{International Conference on Learning Representations (ICLR)}},
year = {2022}
}

@inproceedings{LLWLHL21,
author = {Yuezun Li and Yiming Li and Baoyuan Wu and Longkang Li and Ran He and Siwei Lyu},
title = {{Invisible Backdoor Attack with Sample-Specific Triggers}},
booktitle = {{IEEE International Conference on Computer Vision (ICCV)}},
pages = {16443-16452},
publisher = {IEEE},
year = {2021}
}

@inproceedings{NT21,
author = {Tuan Anh Nguyen and Anh Tuan Tran},
title = {{WaNet - Imperceptible Warping-based Backdoor Attack}},
booktitle = {{International Conference on Learning Representations (ICLR)}},
year = {2021}
}

@inproceedings{DZLLW22,
author = {Wei Du and Yichun Zhao and Boqun Li and Gongshen Liu and Shilin Wang},
title = {{PPT: Backdoor Attacks on Pre-trained Models via Poisoned Prompt Tuning}},
booktitle = {{International Joint Conferences on Artifical Intelligence (IJCAI)}},
pages = {680-686},
publisher = {IJCAI},
year = {2022}
}

@inproceedings{QLCZLWS21,
author = {Fanchao Qi and Mukai Li and Yangyi Chen and Zhengyan Zhang and Zhiyuan Liu and Yasheng Wang and Maosong Sun},
title = {{Hidden Killer: Invisible Textual Backdoor Attacks with Syntactic Trigger}},
booktitle = {{Annual Meeting of the Association for Computational Linguistics and International Joint Conference on Natural Language Processing (ACL/IJCNLP)}},
pages = {443-453},
publisher = {ACL},
year = {2021}
}

@inproceedings{QCLYLS21,
author = {Fanchao Qi and Yangyi Chen and Mukai Li and Yuan Yao and Zhiyuan Liu and Maosong Sun},
title = {{{ONION:} {A} Simple and Effective Defense Against Textual Backdoor Attacks}},
booktitle = {{Conference on Empirical Methods in Natural Language Processing (EMNLP)}},
pages = {9558-9566},
publisher = {ACL},
year = {2021}
}

@inproceedings{WNB18,
author = {Adina Williams and Nikita Nangia and Samuel R. Bowman},
title = {{A Broad-Coverage Challenge Corpus for Sentence Understanding through Inference}},
booktitle = {{Conference of the North American Chapter of the Association for Computational Linguistics: Human Language Technologies (NAACL-HLT)}},
pages = {1112-1122},
publisher = {ACL},
year = {2018}
}

@inproceedings{HBBZMSS21,
author = {Dan Hendrycks and Collin Burns and Steven Basart and Andy Zou and Mantas Mazeika and Dawn Song and Jacob Steinhardt},
title = {{Measuring Massive Multitask Language Understanding}},
booktitle = {{International Conference on Learning Representations (ICLR)}},
year = {2021}
}

@inproceedings{HSWALWWC22,
author = {Edward J. Hu and Yelong Shen and Phillip Wallis and Zeyuan Allen{-}Zhu and Yuanzhi Li and Shean Wang and Lu Wang and Weizhu Chen},
title = {{LoRA: Low-Rank Adaptation of Large Language Models}},
booktitle = {{International Conference on Learning Representations (ICLR)}},
year = {2022}
}

@inproceedings{PZSZY22,
author = {Xudong Pan and Mi Zhang and Beina Sheng and Jiaming Zhu and Min Yang},
title = {{Hidden Trigger Backdoor Attack on {NLP} Models via Linguistic Style Manipulation}},
booktitle = {{USENIX Security Symposium (USENIX Security)}},
pages = {3611-3628},
publisher = {USENIX},
year = {2022}
}

@inproceedings{MLWZM23,
author = {Kai Mei and Zheng Li and Zhenting Wang and Yang Zhang and Shiqing Ma},
title = {{{NOTABLE:} Transferable Backdoor Attacks Against Prompt-based {NLP} Models}},
booktitle = {{Annual Meeting of the Association for Computational Linguistics (ACL)}},
pages = {15551-15565},
publisher = {ACL},
year = {2023}
}

@inproceedings{HZXHYC23,
author = {Yujin Huang and Terry Yue Zhuo and Qiongkai Xu and Han Hu and Xingliang Yuan and Chunyang Chen},
title = {{Training-free Lexical Backdoor Attacks on Language Models}},
booktitle = {{The Web Conference (WWW)}},
pages = {2198-2208},
publisher = {ACM},
year = {2023}
}

@inproceedings{SSN12,
author = {Martin Sundermeyer and Ralf Schl{\"{u}}ter and Hermann Ney},
title = {{{LSTM} Neural Networks for Language Modeling}},
booktitle = {{Conference of the International Speech Communication Association (INTERSPEECH)}},
pages = {194-197},
publisher = {ISCA},
year = {2012}
}

@inproceedings{ZLWJZBSZ24,
author = {Rui Zhang and Hongwei Li and Rui Wen and Wenbo Jiang and Yuan Zhang and Michael Backes and Yun Shen and Yang Zhang},
title = {{Instruction Backdoor Attacks Against Customized LLMs}},
booktitle = {{USENIX Security Symposium (USENIX Security)}},
publisher = {USENIX},
year = {2024}
}

@inproceedings{KKRATSBNSNESGDMSNM23,
author = {Andreas K{\"{o}}pf and Yannic Kilcher and Dimitri von R{\"{u}}tte and Sotiris Anagnostidis and Zhi Rui Tam and Keith Stevens and Abdullah Barhoum and Duc Nguyen and Oliver Stanley and Rich{\'{a}}rd Nagyfi and Shahul ES and Sameer Suri and David Glushkov and Arnav Dantuluri and Andrew Maguire and Christoph Schuhmann and Huu Nguyen and Alexander Mattick},
title = {{OpenAssistant Conversations - Democratizing Large Language Model Alignment}},
booktitle = {{Annual Conference on Neural Information Processing Systems (NeurIPS)}},
publisher = {NeurIPS},
year = {2023}
}

@inproceedings{LWYMWCC24,
author = {Shih{-}Yang Liu and Chien{-}Yi Wang and Hongxu Yin and Pavlo Molchanov and Yu{-}Chiang Frank Wang and Kwang{-}Ting Cheng and Min{-}Hung Chen},
title = {{DoRA: Weight-Decomposed Low-Rank Adaptation}},
booktitle = {{International Conference on Machine Learning (ICML)}},
publisher = {PMLR},
year = {2024}
}

@inproceedings{ZLZLYH24,
author = {Xukun Zhou and Jiwei Li and Tianwei Zhang and Lingjuan Lyu and Muqiao Yang and Jun He},
title = {{Backdoor Attacks with Input-Unique Triggers in {NLP}}},
booktitle = {{European Conference on Machine Learning and Principles and Practice of Knowledge Discovery in Databases (ECML/PKDD)}},
pages = {296-312},
publisher = {Springer},
year = {2024}
}

@inproceedings{CZP25,
author = {Zhuowei Chen and Qiannan Zhang and Shichao Pei},
title = {{Injecting Universal Jailbreak Backdoors into LLMs in Minutes}},
booktitle = {{International Conference on Learning Representations (ICLR)}},
year = {2025}
}

@inproceedings{LLCZLWZL24,
author = {Yanzhou Li and Tianlin Li and Kangjie Chen and Jian Zhang and Shangqing Liu and Wenhan Wang and Tianwei Zhang and Yang Liu},
title = {{BadEdit: Backdooring Large Language Models by Model Editing}},
booktitle = {{International Conference on Learning Representations (ICLR)}},
year = {2024}
}

@inproceedings{PSL22,
author = {Ofir Press and Noah A. Smith and Mike Lewis},
title = {{Train Short, Test Long: Attention with Linear Biases Enables Input Length Extrapolation}},
booktitle = {{International Conference on Learning Representations (ICLR)}},
year = {2022}
}

@inproceedings{LSLZMQ21,
author = {Linyang Li and Demin Song and Xiaonan Li and Jiehang Zeng and Ruotian Ma and Xipeng Qiu},
title = {{Backdoor Attacks on Pre-trained Models by Layerwise Weight Poisoning}},
booktitle = {{Conference on Empirical Methods in Natural Language Processing (EMNLP)}},
pages = {3023-3032},
publisher = {ACL},
year = {2021}
}

@inproceedings{LZWWSZ25,
author = {Chenhao Lin and Chenyang Zhao and Shiwei Wang and Longtian Wang and Chao Shen and Zhengyu Zhao},
title = {{Revisiting Training-Inference Trigger Intensity in Backdoor Attacks}},
booktitle = {{USENIX Security Symposium (USENIX Security)}},
pages = {6359-6378},
publisher = {USENIX},
year = {2025}
}

@article{DCL19,
author = {Jiazhu Dai and Chuanshuai Chen and Yufeng Li},
title = {{A Backdoor Attack Against LSTM-Based Text Classification Systems}},
journal = {{IEEE Access}},
publisher = {IEEE},
year = {2019}
}

@article{SALPBL24,
author = {Jianlin Su and Murtadha H. M. Ahmed and Yu Lu and Shengfeng Pan and Wen Bo and Yunfeng Liu},
title = {{RoFormer: Enhanced transformer with Rotary Position Embedding}},
journal = {{Neurocomputing}},
publisher = {Elsevier},
year = {2024}
}

@article{ZJGGXWFFPL25,
author = {Shuai Zhao and Meihuizi Jia and Zhongliang Guo and Leilei Gan and Xiaoyu Xu and Xiaobao Wu and Jie Fu and Yichao Feng and Fengjun Pan and Anh Tuan Luu},
title = {{A Survey of Recent Backdoor Attacks and Defenses in Large Language Models}},
journal = {{Transactions of Machine Learning Research}},
year = {2025}
}

@article{GDG17,
author = {Tianyu Gu and Brendan Dolan-Gavitt and Siddharth Garg},
title = {{Badnets: Identifying Vulnerabilities in the Machine Learning Model Supply Chain}},
journal = {{CoRR abs/1708.06733}},
year = {2017}
}

@article{LWJLX20,
author = {Yiming Li and Baoyuan Wu and Yong Jiang and Zhifeng Li and Shu{-}Tao Xia},
title = {{Backdoor Learning: {A} Survey}},
journal = {{CoRR abs/2007.08745}},
year = {2020}
}

@article{CLLLS17,
author = {Xinyun Chen and Chang Liu and Bo Li and Kimberly Lu and Dawn Song},
title = {{Targeted Backdoor Attacks on Deep Learning Systems Using Data Poisoning}},
journal = {{CoRR abs/1712.05526}},
year = {2017}
}

@article{O23,
author = {OpenAI},
title = {{{GPT-4} Technical Report}},
journal = {{CoRR abs/2303.08774}},
year = {2023}
}

@article{TLIMLLRGHARJGL23,
author = {Hugo Touvron and Thibaut Lavril and Gautier Izacard and Xavier Martinet and Marie{-}Anne Lachaux and Timoth{\'{e}}e Lacroix and Baptiste Rozi{\`{e}}re and Naman Goyal and Eric Hambro and Faisal Azhar and Aur{\'{e}}lien Rodriguez and Armand Joulin and Edouard Grave and Guillaume Lample},
title = {{LLaMA: Open and Efficient Foundation Language Models}},
journal = {{CoRR abs/2302.13971}},
year = {2023}
}

@article{HZBSZ23,
author = {Hai Huang and Zhengyu Zhao and Michael Backes and Yun Shen and Yang Zhang},
title = {{Composite Backdoor Attacks Against Large Language Models}},
journal = {{CoRR abs/2310.07676}},
year = {2023}
}

@article{T25,
author = {Gemma Team},
title = {{Gemma 3 Technical Report}},
journal = {{CoRR abs/2503.19786}},
year = {2025}
}

@article{T252,
author = {Qwen Team},
title = {{Qwen3 Technical Report}},
journal = {{CoRR abs/2505.09388}},
year = {2025}
}

@article{T253,
author = {Phi Team},
title = {{Phi-4-Mini Technical Report: Compact yet Powerful Multimodal Language Models via Mixture-of-LoRAs}},
journal = {{CoRR abs/2503.01743}},
year = {2025}
}

@article{O25,
author = {Team Olmo},
title = {{Olmo 3}},
journal = {{CoRR abs/2512.13961}},
year = {2025}
}

@article{KMMWARMKAYR20,
author = {Narine Kokhlikyan and Vivek Miglani and Miguel Martin and Edward Wang and Bilal Alsallakh and Jonathan Reynolds and Alexander Melnikov and Natalia Kliushkina and Carlos Araya and Siqi Yan and Orion Reblitz{-}Richardson},
title = {{Captum: {A} unified and generic model interpretability library for PyTorch}},
journal = {{CoRR abs/2009.07896}},
year = {2020}
}

@article{S19,
author = {Robin M. Schmidt},
title = {{Recurrent Neural Networks (RNNs): {A} gentle Introduction and Overview}},
journal = {{CoRR abs/1912.05911}},
year = {2019}
}

@article{HYCLHO24,
author = {Yang Hou and Qiuling Yue and Lujia Chai and Guozhao Liao and Wenbao Han and Wei Ou},
title = {{Double Landmines: Invisible Textual Backdoor Attacks based on Dual-Trigger}},
journal = {{CoRR abs/2412.17531}},
year = {2024}
}

@misc{claude,
author = {Claude Team},
titile = {Claude},
year = 2025,
howpublished = {\url{https://claude.ai/}},
}

@misc{CodeAlpaca,
  author = {Sahil Chaudhary},
  title = {Code Alpaca: An Instruction-following LLaMA model for code generation},
  year = {2023},
  publisher = {GitHub},
  journal = {GitHub repository},
  howpublished = {\url{https://github.com/sahil280114/codealpaca}},
}
